\documentclass[11pt,draftcls,onecolumn]{IEEEtran}
\usepackage{hyperref}

\usepackage{cite}

\ifCLASSINFOpdf
  \usepackage[pdftex]{graphicx}
  \DeclareGraphicsExtensions{.pdf,.jpeg,.png}
\else
  \usepackage[dvips]{graphicx}
  \DeclareGraphicsExtensions{.eps}
\fi
\usepackage[cmex10]{amsmath}

\newtheorem{lemma}{Lemma}
\newtheorem{theorem}{Theorem}

\newtheorem{definition}{Definition}

\usepackage{paralist}
\usepackage{amssymb}
\usepackage{amsfonts}
\newcommand{\tabincell}[2]{\begin{tabular}{@{}#1@{}}#2\end{tabular}}

\begin{document}
%
\title{Distributed Multilevel Diversity Coding}
%
%
%

\author{Zhiqing Xiao, Jun Chen, Yunzhou Li, and Jing Wang
\thanks{Zhiqing Xiao is with the Department of Electronic Engineering, Tsinghua University, Beijing, 100084, P. R. China (email: xzq.xiaozhiqing@gmail.com).}
\thanks{Jun Chen is with the Department
of Electrical and Computer Engineering, McMaster University,
Hamilton, ON L8S 4K1, Canada  (email: junchen@ece.mcmaster.ca).}
\thanks{Yunzhou Li and Jing Wang are with Research Institute of Information Technology, Tsinghua University, Beijing, 100084, P. R. China (email:\{liyunzhou,wangj\}@tsinghua.edu.cn).}%
}

\maketitle

\begin{abstract}
In distributed multilevel diversity coding, $K$ correlated sources (each with $K$ components) are encoded in a distributed manner such that, given the outputs from any $\alpha$ encoders, the decoder can reconstruct the first $\alpha$ components of each of the corresponding $\alpha$ sources. For this problem, the optimality of a multilayer Slepian-Wolf coding scheme based on binning and superposition is established when $K\leq 3$. The same conclusion is shown to hold for general $K$ under a certain symmetry condition, which generalizes a celebrated result by Yeung and Zhang.
\end{abstract}


%
\IEEEpeerreviewmaketitle

\section{Introduction}
\label{sec:introduce}
%
%
%
%

Consider the scenario where $K$ correlated sources $U_1,U_2,\cdots,U_K$ are compressed by $K$ sensors in a distributed manner and then forwarded to a fusion center for joint reconstruction. This is exactly the classical distributed data compression problem, for which Slepian-Wolf coding \cite{slepian1973,cover1975} is known to be rate-optimal. However, to attain this best compression efficiency, encoding at each sensor is performed under the assumption that all the other sensors are functioning properly; as a consequence, inactivity of one or more sensors typically leads to a complete decoding failure at the fusion center. Alternatively, each sensor can compress its observation using conventional point-to-point data compression methods without capitalizing the correlation among different sources so that the maximum system robustness can be achieved. In view of these two extreme cases, a natural question arises whether there exists a tradeoff between compression efficiency and system robustness in distributed data compression.

One approach to realize this tradeoff is as follows. Specifically, we decompose each $U_k$ into $K$ components $U_{k,1},U_{k,2},\cdots,U_{k,K}$, ordered according to their importance, and encode them in such a way that, given the outputs from any $\alpha$ sensors, the fusion center can reconstruct the first $\alpha$ components of each of the corresponding $\alpha$ sources. The aforedescribed two extreme cases correspond to $(U_{k,1},\cdots,U_{k,K-1},U_{k,K})=(0,\cdots,0,U_{k})$ and $(U_{k,1},U_{k,2},\cdots,U_{k,K})=(U_k,0,\cdots,0)$, respectively. One can realize a flexible tradeoff between compression efficiency and system robustness by adjusting the amount of information allocated to different components. We shall refer to this problem as distributed multilevel diversity coding (D-MLDC) since it reduces to the well-known (symmetrical) multilevel diversity coding (MLDC)  problem when $U_{1,\alpha}=U_{2,\alpha}=\cdots=U_{K,\alpha}$ almost surely for all $\alpha$.

The concept of MLDC was introduced by Roche~\cite{roche1992} and more formally by Yeung~\cite{yeung1995} though research on diversity coding can be traced back to Singleton's work on maximum distance separable codes~\cite{singleton1964}. The symmetric version of this problem has received particular attention~\cite{roche1997}, and arguably the  culminating achievement of this line of research is the complete characterization of the admissible rate region of symmetrical MLDC by Yeung and Zhang~\cite{yeung1999}. Some recent developments related to MLDC can be found in~\cite{mohajer2010, balasubramanian2013, jiang2014, tian2015}.

The goal of the present paper to characterize the performance limits of D-MLDC, which, we hope, may provide some useful insights into the tradeoff between compression efficiency and system robustness in distributed data compression. More fundamentally, we aim to examine the principle of superposition~\cite{yeung1995} in the context of D-MLDC. Although superposition (or more generally, layering) is a common way to construct sophisticated schemes based on simple building blocks and often yields the best known achievability results, establishing the optimality of such constructions is rarely straightforward, especially when encoding is performed in a distributed manner. In fact, even for the centralized encoding setup studied in \cite{yeung1999}, the proof of the optimality of superposition is already highly non-trivial. This difficulty can be partly attributed to the fact that it is often a technically formidable task to extract layers from a generic scheme using information inequalities in a converse argument, even in cases where the use of layered constructions may appear rather natural.

From this perspective, our work can be viewed as an initial step towards a better understanding of layered schemes for distributed compression of correlated sources. We shall propose a multilayer Slepian-Wolf coding scheme based on binning and superposition, and establish its optimality for D-MLDC when $K\leq 3$. This scheme is also shown to be optimal for general $K$ under a certain symmetry condition, which generalizes the aforementioned result by Yeung and Zhang on symmetrical MLDC \cite{yeung1999}. The main technical difficulty encountered in our proof is that it appears to be infeasible to characterize the admissible rate region of D-MLDC by deriving inner and outer bounds separately and then making a direct comparison based on their explicit expressions. To circumvent this difficulty, we follow the approach in \cite{yeung1999}, where the analysis of the inner bound and that of the outer bound are conceptually intertwined. Specifically, we analyze certain linear programs associated with the achievable rate region of the proposed scheme and leverage the induced Lagrange multipliers to establish the entropy inequalities that are needed for a matching converse. Since the problem considered here is more general than that in \cite{yeung1999}, the relevant linear programs and entropy inequalities are inevitably more sophisticated. It is worth mentioning that, in a broad sense, the strategy of determining an information-theoretic limit by connecting achievability and converse results to a common optimization problem (not necessarily linear) via duality has find applications far beyond MLDC (see, e.g., \cite{song2012}).

The rest of this paper is organized as follows. We state the basic definitions and the main results in Section \ref{sec:formula}. Section~\ref{sec:prove} contains a high-level description of our general approach. The detailed proofs can be found in Sections~\ref{sec:main3} and \ref{sec:prove_sym}. We conclude the paper in Section~\ref{sec:conclude}.


\emph{Notation:}
 Random vectors $\left( {{X}_{t}},t\in T \right)$ and $\left( {{X}_{t,t'}},t\in T, t'\in T' \right)$ are sometimes abbreviated as  ${{X}_{T}}$ and ${{X}_{T,T'}}$, respectively. For two integers ${{x}_{1}},{{x}_{2}}\in \mathbb{Z}$, we define $\left[ {{x}_{1}}:{{x}_{2}} \right]\triangleq\left\{ x\in \mathbb{Z}:{{x}_{1}}\le x\le {{x}_{2}} \right\}$. The cardinality of a finite set $V$ is denoted by $|V|$; moreover, for any $T\subseteq \left[ 1:\left| V \right| \right]$, let ${{\left\langle V \right\rangle }_{T}}\triangleq\left\{ v\in V:\left| \left\{ {v'}\in V:{v'}\le v \right\} \right|\in T \right\}$. We often do not distinguish between a singleton and its element.





\section{Problem Formulation and Main Results}
\label{sec:formula}

\subsection{System Model}
\label{sec:model}

Let $U_{k,\left[1:K\right]}$, $k\in\left[1:K\right]$, be $K$ vector sources. We assume\footnote{This assumption can be relaxed to a certain extent and can be modified in various ways. In this paper we do not seek to present our results in their most general forms since the resulting statements and expressions may become rather unwieldy.} that $U_{\left[1:K\right],\alpha}$, $\alpha\in\left[1:K\right]$, are mutually independent whereas, for each $\alpha$, the components of $U_{\left[1:K\right],\alpha}$ (i.e., $U_{k,\alpha}$, $k\in\left[1:K\right]$) can be arbitrarily correlated. Let $\left\{U_{\left[1:K\right],\left[1:K\right]}(t)\right\}_{t=1}^\infty$ be i.i.d. copies of $U_{\left[1:K\right],\left[1:K\right]}$.

\begin{figure}[thbp!]
	\centering
		\includegraphics{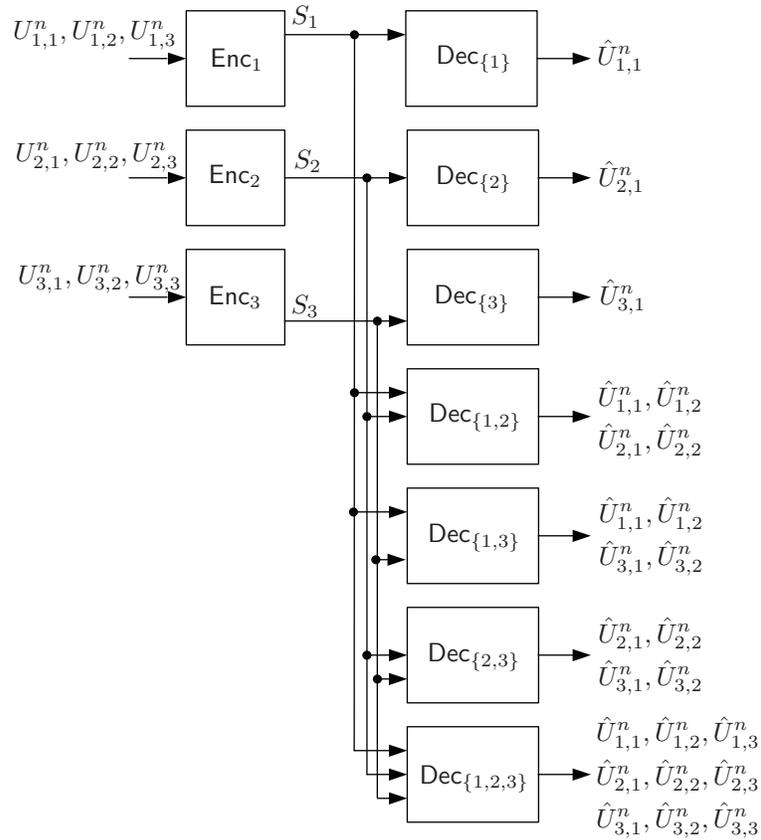}
	\caption{System diagram for D-MLDC with $K=3$.}
	\label{fig:systemdiagram}
\end{figure}

An $\left( n,\left( {{M}_{k}},k\in \left[ 1:K \right] \right) \right)$ D-MLDC system consists of:
\begin{enumerate}[$\bullet$]
\item $K$ encoders, where encoder ${\mathsf{Enc}_{k}}$ ($k\in \left[ 1:K \right]$) maps the source sequence $U_{k,\left[ 1:K \right]}^{n}$ to a symbol ${{S}_{k}}$ in $\left[ 1:{{M}_{k}} \right]$, i.e.,
\begin{align*}
{\mathsf{Enc}_{k}}:\prod\limits_{\alpha =1}^{K}{\mathcal{U}_{k,\alpha }^{n}} & \to  \left[ 1:{{M}_{k}} \right], \\
U_{k,\left[ 1:K \right]}^{n} & \mapsto  \quad {{S}_{k}},
\end{align*}
\item ${{2}^{K}}-1$ decoders, where decoder ${\mathsf{Dec}_{V}}$ ($\emptyset \subsetneqq V \subseteq \left[ 1:K \right]$) produces a reconstruction of $U_{V,\left[ 1:\left| V \right| \right]}^{n}$, denoted by $\hat{U}_{V,\left[ 1:\left| V \right| \right]}^{n}$, based on ${{S}_{V}}$, i.e.,
\begin{align*}
{\mathsf{Dec}_{V}}:\prod\limits_{k\in V}{\left[ 1:{{M}_{k}} \right]} &\to  \prod\limits_{k\in V}{\prod\limits_{\alpha =1}^{\left| V \right|}{\mathcal{U}_{k,\alpha }^{n}}}, \\
{{S}_{V}} \quad & \mapsto  \hat{U}_{V,\left[ 1:\left| V \right| \right]}^{n}.
\end{align*}
\end{enumerate}
A D-MLDC system with $K=3$ is illustrated in Fig.~\ref{fig:systemdiagram}.

\subsection{Admissible Rate Region}
\label{sec:admit}

A rate tuple $\left( {{R}_{k}},k\in \left[ 1:K \right] \right)$ is said to admissible if, for any $\epsilon >0$, there exists an $\left( n,\left( {{M}_{k}},k\in \left[ 1:K \right] \right) \right)$ D-MLDC system such that

(1) (Rate Constraints)
\begin{align}
\label{eq:constraint1}
\frac{1}{n}\log {{M}_{k}} \leq {{R}_{k}}+\epsilon , \quad k\in \left[ 1:K \right],
\end{align}

(2) (Reconstruction Constraints)
\begin{align}
\label{eq:constraint2}
\Pr \left\{ U_{V,\left[ 1:\left| V \right| \right]}^{n}\ne \hat{U}_{V,\left[ 1:\left| V \right| \right]}^{n} \right\} \leq \epsilon , \quad \emptyset \ne V\subseteq \left[ 1:K \right].
\end{align}

The admissible rate region $\mathcal{R}^*_K$ is defined as the set of all admissible rate tuples.

\subsection{Multilayer Slepian-Wolf Coding}
\label{sec:code}

We shall propose a D-MLDC scheme, which can be viewed as a natural extension of that in~\cite{yeung1999} to the distributed encoding setup. This scheme, termed multilayer Slepian-Wolf coding, includes two steps: intralayer coding and interlayer coding.

\begin{enumerate}[$\bullet$]
\item \emph{Intralayer Coding:} For each $\alpha\in\left[1:K\right]$, encoder $k$ ($k\in \left[1:K \right]$) compresses $U^n_{k,\alpha}$ using the conventional binning scheme\footnote{Here one can in fact use universal Slepian-Wolf coding so that encoding and decoding can be performed without the knowledge of the source distribution \cite{csiszar1982}.} of rate $r_{k,\alpha}$;  correct reconstruction of $U^n_{k,\alpha}$, $\alpha\in V$, based on the corresponding bin indices is ensured (with high probability) for all $V\subseteq[1:K]$ with $|V|=\alpha$ if $\left( {{r}_{k,\alpha }},k\in \left[ 1:K \right] \right)\in{{\mathcal{R}}_{K,\alpha }}$, where
\begin{align*}
 &{{\mathcal{R}}_{K,\alpha }} \triangleq \left\{ \left( {{r}_{k,\alpha }},k\in \left[ 1:K \right] \right): \sum\limits_{k\in V}{{{r}_{k,\alpha }}}\ge H\left( {{U}_{V,\alpha }}|{{U}_{{V'},\alpha }} \right),\quad V\in {{\mathbb{V}}_{K,\alpha }},V'\in {\mathbb{V}'_{K,\alpha }}\left[ V \right]  \right\}
\end{align*}
with
\begin{align*}
&{{\mathbb{V}}_{K,\alpha }} \triangleq\left\{ V\subseteq \left[ 1:K \right]:1\le \left| V \right|\le \alpha  \right\}, \\
&{\mathbb{V}'_{K,\alpha }}\left[ V \right]  \triangleq  \left\{ {V'}\subseteq \left[ 1:K \right]\backslash V:\left| {{V'}} \right|+\left| V \right|=\alpha  \right\},\quad V\in {{\mathbb{V}}_{K,\alpha }}.
\end{align*}
\item \emph{Interlayer Coding:} In this step, encoder $k$ ($k\in \left[1:K \right]$) generates its output by combining the bin indices associated with $U^n_{k,\alpha}$, $\alpha\in\left[1:K\right]$, via superposition. Note that the resulting rate region
    \begin{align*}
{{\mathcal{R}}_{K}}\triangleq\left\{\sum\limits_{\alpha=1}^K\left( {{r}_{k,\alpha }},k\in \left[ 1:K \right] \right): \left( {{r}_{k,\alpha }},k\in \left[ 1:K \right] \right)\in{{\mathcal{R}}_{K,\alpha }},\quad\alpha\in \left[ 1:K \right] \right\}
\end{align*}
is an inner bound of $\mathcal{R}^*_K$, i.e.,
\begin{align}
{{\mathcal{R}}_{K}}\subseteq\mathcal{R}^*_K.\label{eq:innerbound}
\end{align}
\end{enumerate}




\subsection{Main Results}
\label{sec:main}

Our first main result shows that ${{\mathcal{R}}_{K}}$ coincides with $\mathcal{R}^*_K$ when $K\le 3$.

\begin{theorem}
\label{th:main3}
$\mathcal{R}^*_K={{\mathcal{R}}_{K}}$ for $K\le 3$.
\end{theorem}

To state our second main result, we need the following definition.
\begin{definition}[Symmetrical Source]
\label{def:sym}
We say that the distribution of ${{U}_{\left[ 1:K \right],\left[ 1:K \right]}}$ is symmetrical entropy-wise if $H\left( {{U}_{V,\alpha }} \right)=H\left( {{U}_{{V'},\alpha }} \right)$ for all $\alpha \in \left[ 1:K \right]$ and $V,{V'}\subseteq \left[ 1:K \right]$ with $\left| V \right|=\left| {{V'}} \right|$.
\end{definition}

It is worth noting that the symmetrical MLDC problem studied in \cite{yeung1999} corresponds to the special case where $H\left( {{U}_{V,\alpha }} \right)=H\left( {{U}_{{V'},\alpha }} \right)$ for all $\alpha \in \left[ 1:K \right]$ and $\emptyset \subsetneqq V,{V'}\subseteq \left[ 1:K \right]$.

\begin{theorem}
\label{th:main_sym}
If the distribution of ${{U}_{\left[ 1:K \right],\left[ 1:K \right]}}$ is symmetrical entropy-wise, then $\mathcal{R}^*_{K}={{\mathcal{R}}_{K}}$.
\end{theorem}

\section{Outline of a General Approach}
\label{sec:prove}

In this section we attempt to give an outline of our general approach, which is in principle not restricted to the cases covered by Theorems \ref{th:main3} and \ref{th:main_sym}. On a conceptual level, this approach was originated in \cite{yeung1999} (and made more evident in \cite{jiang2014}). It consists of three major steps:
\begin{enumerate}
\item characterize the supporting hyperplanes of ${{\mathcal{R}}_{K}}$ (more precisely, the supporting hyperplanes of ${{\mathcal{R}}_{K,\alpha }}$, $\alpha\in\left[1:K\right]$) via the analysis of the corresponding linear programs;

\item establish a class of entropy inequalities based on the Lagrange multipliers induced by the aforementioned linear programs;

\item derive a tight outer bound on ${{\mathcal{R}}_{K}}$ by leveraging these entropy inequalities.
\end{enumerate}


\subsection{Linear Program}
\label{sec:r}



Each supporting hyperplane of $\mathcal{R}_{K,\alpha}$ is associated with a linear program
\begin{align*}
{\mathsf{LP}^{\mathbf{w}}_{K,\alpha }}: \min \quad &
\sum\limits_{k=1}^{K}{{{w}_{k}}{{r}_{k,\alpha }}} \\
\operatorname{over} \quad &
{{r}_{k,\alpha }}, \quad k\in \left[ 1:K \right], \\
\operatorname{s.t.} \quad &
\sum\limits_{k\in V}{{{r}_{k,\alpha }}} \ge H\left( {{U}_{V,\alpha }}|{{U}_{{V'},\alpha }} \right),\quad V\in {{\mathbb{V}}_{K,\alpha }},V'\in {\mathbb{V}'_{K,\alpha }}\left[ V \right],
\end{align*}
where $\mathbf{w}\triangleq\left( {{w}_{k}},k\in \left[ 1:K \right] \right)\in \mathbb{R}_{+}^{K}$.
It often suffices to consider the case where the weights $ {{w}_{k}}$, $k\in \left[ 1:K \right]$, are ordered. For this reason, we define
\begin{align*}
{\mathbb{\mathbb{W} }_{K}}\triangleq\left\{\mathbf{w}:{{w}_{1}}\ge {{w}_{2}}\ge \cdots \ge {{w}_{K}}\geq{{w}_{K+1}}\triangleq 0 \right\}.
\end{align*}
Moreover, to facilitate subsequent analysis, we introduce the following partition\footnote{For $\mathbf{w}\in {{\mathbb{W}}_{K}}$, we have ${{w}_{l}}>\frac{1}{\alpha -l}\sum_{k=l+1}^{K}{{{w}_{k}}} \Rightarrow {{w}_{l'}}>\frac{1}{\alpha -l'}\sum_{k=l'+1}^{K}{{{w}_{k}}}$, $l'\in\left[1:l\right]$, and ${{w}_{l}}\le \frac{1}{\alpha -l}\sum_{k=l+1}^{K}{{{w}_{k}}}\Rightarrow{{w}_{l'}}\le \frac{1}{\alpha -l'}\sum_{k=l'+1}^{K}{{{w}_{k}}}$, $l'\in\left[l:\alpha-1\right]$. Therefore, ${{\mathbb{W}}^{\left( 0 \right)}_{K,\alpha }},\cdots,{{\mathbb{W}}^{\left( \alpha-1 \right)}_{K,\alpha }}$ indeed form a partition of $\mathbb{W}_{K}$.} of ${\mathbb{\mathbb{W} }_{K}}$ for each $\alpha\in\left[2:K\right]$:
\begin{align*}
&{{\mathbb{W}}^{\left( 0 \right)}_{K,\alpha }}
\triangleq  \left\{ \mathbf{w}\in {{\mathbb{W}}_{K}}:{{w}_{1}}\le \frac{1}{\alpha -1}\sum\limits_{k=2}^{K}{{{w}_{k}}} \right\} , \\
&{{\mathbb{W}}^{\left( l \right)}_{K,\alpha }} \triangleq \left\{ \mathbf{w}\in {{\mathbb{W}}_{K}}:{{w}_{l}}>\frac{1}{\alpha -l}\sum\limits_{k=l+1}^{K}{{{w}_{k}}}  \text{ and }{{w}_{l+1}}\le \frac{1}{\alpha -\left( l+1 \right)}\sum\limits_{k=l+2}^{K}{{{w}_{k}}} \right\}, \quad l \in \left[1 : \alpha -2 \right],  \\
&{{\mathbb{W}}^{\left( \alpha-1 \right)}_{K,\alpha }} \triangleq
   \left\{ \mathbf{w}\in {{\mathbb{W}}_{K}}:{{w}_{\alpha -1}}>\sum\limits_{k=\alpha }^{K}{{{w}_{k}}} \right\}.
\end{align*}
We set ${{\mathbb{W}}^{\left( 0 \right)}_{K,1}}\triangleq{\mathbb{\mathbb{W} }_{K}}$.


\begin{definition}[Lagrange Multiplier]
\label{def:lagrange}
 We say\footnote{It can be shown that $\left( {{c}_{V|{V}',\alpha }},V\in {{\mathbb{V}}_{K,\alpha }},{V}'\in {{\mathbb{V}}'_{K,\alpha }}\left[ V \right] \right)$ is an optimal Lagrange multiplier of ${\mathsf{LP}^{\mathbf{w}}_{K,\alpha }}$ if and only if it is an optimal solution to the (asymmetric) dual problem of ${\mathsf{LP}^{\mathbf{w}}_{K,\alpha }}$.} $\left( {{c}_{V|{V}',\alpha }},V\in {{\mathbb{V}}_{K,\alpha }},{V}'\in {{\mathbb{V}}'_{K,\alpha }}\left[ V \right] \right)$ is an optimal Lagrange multiplier of ${\mathsf{LP}^{\mathbf{w}}_{K,\alpha }}$ with $\mathbf{w}\in \mathbb{R}_{+}^{K}$ if
\begin{align}
&\sum\limits_{V\in {{\mathbb{V}}_{K,\alpha }}}{\sum\limits_{{V}'\in {{\mathbb{V}}'_{K,\alpha }}\left[ V \right]}{{{c}_{V|{V}',\alpha }}H\left( {{U}_{V,\alpha }}|{{U}_{{V}',\alpha }} \right)}}= {{f}^{\mathbf{w}}_{{{\alpha }}}},\label{eq:opt} \\
&\sum\limits_{V\in {{\mathbb{V}}_{K,\alpha }}:k\in V}{\sum\limits_{{V}'\in {{\mathbb{V}}'_{K,\alpha }}\left[ V \right]}{{{c}_{V|{V}',\alpha }}}} = {{w}_{k}},\quad k\in \left[ 1:K \right], \label{eq:connection}\\
&{{c}_{V|{V}',\alpha }} \ge 0,\quad V\in {{\mathbb{V}}_{K,\alpha }},{V}'\in {{\mathbb{V}}'_{K,\alpha }}\left[ V \right],\nonumber
\end{align}
where ${{f}^{\mathbf{w}}_{{{\alpha }}}}$ denotes the optimal value of ${\mathsf{LP}^{\mathbf{w}}_{K,\alpha }}$.
\end{definition}

It is in general not easy to find optimal solution $\left( r_{k,\alpha}^{\operatorname{opt}},k\in \left[ 1:K \right] \right)$ and optimal Lagrange multiplier $\left( {{c}_{V|{V'},\alpha }},V\in {{\mathbb{V}}_{K,\alpha }},{V'}\in {\mathbb{V}'_{K,\alpha }}\left[ V \right] \right)$ for a given $\mathsf{LP}^{\mathbf{w}}_{K,\alpha}$ (see Section \ref{sec:main3} for a detailed analysis of $\mathsf{LP}^{\mathbf{w}}_{3,2}$). However, the task becomes relatively straightforward when $\alpha=1$ or $\alpha=K$ as shown by the following two lemmas (which can be proved via direct verification).

\begin{lemma}
\label{lem:lpk1}
For linear program $\mathsf{LP}^{\mathbf{w}}_{K,1}$ with $\mathbf{w}\in \mathbb{R}_{+}^{K}$, $\left( r_{k,1}^{\operatorname{opt}},k\in \left[ 1:K \right] \right)$ is an optimal solution and $\left( {{c}_{V|\emptyset,1}},V\in {{\mathbb{V}}_{K,1}} \right) $ is the unique optimal Lagrange multiplier, where
\begin{align*}
&r_{k,1}^{\operatorname{opt}}\triangleq H\left( {{U}_{k,1}} \right),\quad k\in \left[ 1:K \right],\\
&{{c}_{\{k\}|\emptyset,1}}\triangleq{{w}_{k}},\quad k\in \left[ 1:K \right].
\end{align*}
\end{lemma}

\begin{lemma}
\label{lem:lpkk}
For linear program $\mathsf{LP}^{\mathbf{w}}_{K,K}$ with $\mathbf{w}\in{\mathbb{\mathbb{W} }_{K}}$, $\left( r_{k,K}^{\operatorname{opt}},k\in \left[ 1:K \right] \right)$ is an optimal solution and $\left( {{c}_{V|[1:K]\backslash V,K}},V\in{{\mathbb{V}}_{K,K}} \right) $ is an optimal Lagrange multiplier, where
\begin{align*}
&r_{k,K}^{\operatorname{opt}}\triangleq H\left( {{U}_{k,K}}|{{U}_{\left[ k+1:K \right],K}} \right),\quad k\in \left[ 1:K \right],\\
&{{c}_{\left[ 1:k \right]|\left[k+1:K\right],K}} \triangleq {{w}_{k}}-{{w}_{k+1}},\quad k\in \left[ 1:K \right], \\
&{{c}_{V|[1:K]\backslash V,K}}\triangleq 0,\quad \text{otherwise}.
\end{align*}
The general case $\mathbf{w}\in \mathbb{R}_{+}^{K}$  can be reduced to the case $\mathbf{w}\in{\mathbb{\mathbb{W} }_{K}}$ via suitable relabelling.
\end{lemma}

\subsection{Entropy Inequality}
\label{sec:ineq}


In this step we aim to establish a class of entropy inequalities needed for a matching converse by exploiting the properties of optimal Lagrange multipliers of $\mathsf{LP}^{\mathbf{w}}_{K,\alpha}$, $\alpha\in\left[1:K\right]$. More precisely, we shall identify suitable conditions under which there exist optimal Lagrange multipliers $\left( {{c}_{V|{V'},\alpha }},V\in {{\mathbb{V}}_{K,\alpha }},{V'}\in {\mathbb{V}'_{K,\alpha }}\left[ V \right] \right)$, $\alpha \in \left[ 1:K \right]$, such that
\begin{align}
\sum\limits_{V\in {{\mathbb{V}}_{K,\alpha'}}}{\sum\limits_{{V'}\in {\mathbb{V}'_{K,\alpha'}}\left[ V \right]}{{{c}_{V|{V'},\alpha'}}H\left( {{X}_{V}}|{{X}_{V'}} \right)}}  \ge \sum\limits_{V\in {{\mathbb{V}}_{K,\alpha }}}{\sum\limits_{{V'}\in {\mathbb{V}'_{K,\alpha }}\left[ V \right]}{{{c}_{V|{V'},\alpha }}H\left( {{X}_{V}}|{{X}_{V'}} \right)}}\label{eq:star}
\end{align}
for all $X_{\left[1:K\right]}$ and $\alpha\geq\alpha'$.


The following lemma indicates that (\ref{eq:star}) always holds when $\alpha'=1$.

\begin{lemma}
\label{lem:ineq2}

Let $\left( {{c}_{V|\emptyset,1}},V\in {{\mathbb{V}}_{K,1}} \right)$ and $\left( {{c}_{V|{V'},\alpha }},V\in {{\mathbb{V}}_{K,\alpha }},{V'}\in {\mathbb{V}'_{K,\alpha }}\left[ V \right] \right)$ be optimal Lagrange multipliers of $\mathsf{LP}^{\mathbf{w}}_{K,1}$ and $\mathsf{LP}^{\mathbf{w}}_{K,\alpha}$, respectively. We have
\begin{align*}
\sum\limits_{V\in {{\mathbb{V}}_{K,1}}}{{{c}_{V|\emptyset,1}}H\left( {{X}_{V}} \right)}\ge \sum\limits_{V\in {{\mathbb{V}}_{K,\alpha}}}{\sum\limits_{{V'}\in {\mathbb{V}'_{K,\alpha}}\left[ V \right]}{{{c}_{V|{V'},\alpha}}H\left( {{X}_{V}}|{{X}_{V'}} \right)}}
\end{align*}
for all $X_{[1:K]}$.
\end{lemma}
\begin{IEEEproof}
According to~(\ref{eq:connection}),
\begin{align}
c_{\{k\}|\emptyset,1}=w_k=\sum\limits_{V\in {{\mathbb{V}}_{K,\alpha }}:k\in V}{\sum\limits_{{V}'\in {{\mathbb{V}}'_{K,\alpha }}\left[ V \right]}{{{c}_{V|{V}',\alpha }}}},\quad k\in\left[1:K\right].\label{eq:identity}
\end{align}
It can be verified that
\begin{align}
\sum\limits_{V\in {{\mathbb{V}}_{K,1}}}{{{c}_{V|\emptyset,1}}H\left( {{X}_{V}} \right)}&=\sum\limits_{k=1}^K{{{c}_{\{k\}|\emptyset,1}}H\left( {{X}_{k}} \right)}\nonumber\\
&=\sum\limits_{k=1}^K\sum\limits_{V\in {{\mathbb{V}}_{K,\alpha }}:k\in V}{\sum\limits_{{V}'\in {{\mathbb{V}}'_{K,\alpha }}\left[ V \right]}{{{c}_{V|{V}',\alpha }}}}H(X_k)\label{eq:invokeid}\\
&=\sum\limits_{V\in {{\mathbb{V}}_{K,\alpha }}}{\sum\limits_{{V}'\in {{\mathbb{V}}'_{K,\alpha }}\left[ V \right]}{{{c}_{V|{V}',\alpha }}}}\sum\limits_{k\in V}H(X_k)\nonumber\\
&\geq\sum\limits_{V\in {{\mathbb{V}}_{K,\alpha }}}{\sum\limits_{{V}'\in {{\mathbb{V}}'_{K,\alpha }}\left[ V \right]}{{{c}_{V|{V}',\alpha }}}}H(X_{V})\nonumber\\
&\geq\sum\limits_{V\in {{\mathbb{V}}_{K,\alpha }}}{\sum\limits_{{V}'\in {{\mathbb{V}}'_{K,\alpha }}\left[ V \right]}{{{c}_{V|{V}',\alpha }}}}H(X_{V}|X_{V'}),\nonumber
\end{align}
where (\ref{eq:invokeid}) is due to (\ref{eq:identity}). This completes the proof of Lemma \ref{lem:ineq2}.
\end{IEEEproof}


\subsection{Outer Bound}

As shown by the following lemma,  the existence of entropy inequalities (\ref{eq:star}) implies that $\mathcal{R}_K$ is an outer bound of $\mathcal{R}^*_K$.



\begin{lemma}
\label{lem:outer}
If for any $\mathbf{w}\in \mathbb{R}_{+}^{K}$, there exist optimal Lagrange multipliers $\left( {{c}_{V|{V'},\alpha }},V\in {{\mathbb{V}}_{K,\alpha }},{V'}\in {\mathbb{V}'_{K,\alpha }}\left[ V \right] \right)$, $\alpha \in \left[ 1:K \right]$, such that (\ref{eq:star}) holds, then 
\begin{align}
\mathcal{R}^*_K\subseteq\mathcal{R}_K.\label{eq:outerbound}
\end{align}
\end{lemma}
\begin{IEEEproof}
Let $\left( {{R}_{k}}, k\in \left[ 1:K \right] \right)$ be an arbitrary admissible rate tuple. It suffices to show that
\begin{align}
\sum\limits_{k=1}^{K}{{{w}_{k}}{{R}_{k}}}\ge \sum\limits_{\alpha =1}^{K}{{{f}^{\mathbf{w}}_{{{\alpha }}}}},\label{eq:supportyhyper}
\end{align}
from which (\ref{eq:outerbound}) follows immediately. We shall prove via induction that, for any D-MLDC system satisfying
(\ref{eq:constraint1}) and (\ref{eq:constraint2}),
\begin{align}
 \sum\limits_{k=1}^{K}{{{w}_{k}}\left(R_k+\epsilon\right)}&\geq \sum\limits_{\alpha =1}^{\beta }{{{f}^{\mathbf{w}}_{{{\alpha }}}}}  +\frac{1}{n}\sum\limits_{V\in {{\mathbb{V}}_{K,\beta }}}{\sum\limits_{{V'}\in {\mathbb{V}'_{K,\beta }}\left[ V \right]}{{{c}_{V|{V'},\beta }}H\left( {{S}_{V}}|U_{\left[ 1:K \right],\left[ 1:\beta  \right]}^{n},{{S}_{V'}} \right)}} \nonumber\\
 &\quad -\beta{{\delta }_{\epsilon }} \sum\limits_{k=1}^{K}{{{w}_{k}}} ,\quad \beta \in \left[ 1:K \right],\label{eq:induction}
\end{align}
where $\delta_{\epsilon}$ tends to zero as $\epsilon\rightarrow 0$. One can deduce (\ref{eq:supportyhyper}) from (\ref{eq:induction}) by setting $\beta=K$ and sending $\epsilon\rightarrow 0$. 

It can be verified that
\begin{align}
\sum\limits_{k=1}^{K}{{{w}_{k}}\left(R_k+\epsilon\right)}&\geq\frac{1}{n}\sum\limits_{k=1}^{K}{{w}_{k}}\log M_k\nonumber\\
&\geq\frac{1}{n}\sum\limits_{k=1}^{K}{{w}_{k}}H(S_k)\nonumber\\
&=\frac{1}{n}\sum\limits_{V\in {{\mathbb{V}}_{K,1}}}{{{c}_{V|\emptyset,1}}H\left( {{S}_{V}} \right)},\label{eq:invokelemma1}
\end{align}
where (\ref{eq:invokelemma1}) is due to the fact (see Lemma \ref{lem:lpk1}) that  the optimal Lagrange multiplier $\left( {{c}_{V|\emptyset,1}},V\in {{\mathbb{V}}_{K,1}} \right) $ is uniquely given by ${{c}_{\{k\}|\emptyset,1}}\triangleq{{w}_{k}}$, $k\in \left[ 1:K \right]$. Note that
\begin{align}
H\left( {{S}_{V}} \right)&\geq H\left(U^n_{V,1}\right)+H\left(S_V|U^n_{V,1}\right)-H\left(U^n_{V,1}|S_V\right)\nonumber\\
&\geq H\left(U^n_{V,1}\right)+H\left(S_V|U^n_{V,1}\right)-n\delta_{\epsilon}\label{eq:Fano}\\
&\geq H\left(U_{V,1}\right)+H\left(S_V|U^n_{\left[1:K\right],1}\right)-n\delta_{\epsilon},\quad V\in {{\mathbb{V}}_{K,1}},\label{eq:sub1}
\end{align}
where (\ref{eq:Fano}) follows by (\ref{eq:constraint2}) and Fano's inequality. Substituting (\ref{eq:sub1}) into (\ref{eq:invokelemma1}) and invoking (\ref{eq:opt}) proves (\ref{eq:induction}) for $\beta=1$. Now assume that (\ref{eq:induction}) holds for $\beta=B-1$. In view of (\ref{eq:star}), we have
\begin{align}
&\sum\limits_{V\in {{\mathbb{V}}_{K,B-1}}}{\sum\limits_{{V'}\in {\mathbb{V}'_{K,B-1}}\left[ V \right]}{{{c}_{V|{V'},B-1}}H\left( {{S}_{V}}|U_{\left[ 1:K \right],\left[ 1:B-1 \right]}^{n},{{S}_{V'}} \right)}}\nonumber \\
  &\ge \sum\limits_{V\in {{\mathbb{V}}_{K,B}}}{\sum\limits_{{V'}\in {\mathbb{V}'_{K,B}}\left[ V \right]}{{{c}_{V|{V'},B}}H\left( {{S}_{V}}|U_{\left[ 1:K \right],\left[ 1:B-1 \right]}^{n},{{S}_{V'}} \right)}}.\label{eq:firstcont}
\end{align}
It can be verified that
\begin{align}
&H\left( {{S}_{V}}|U_{\left[ 1:K \right],\left[ 1:B-1 \right]}^{n},{{S}_{V'}} \right)\nonumber\\
&=H\left( U^n_{V,B},{{S}_{V}}|U_{\left[ 1:K \right],\left[ 1:B-1 \right]}^{n},{{S}_{V'}} \right)-H\left( U^n_{V,B}|U_{\left[ 1:K \right],\left[ 1:B-1 \right]}^{n},{{S}_{V}},{{S}_{V'}} \right)\nonumber\\
&\geq H\left( U^n_{V,B},{{S}_{V}}|U_{\left[ 1:K \right],\left[ 1:B-1 \right]}^{n},{{S}_{V'}} \right)-n\delta_{\epsilon},\quad V\in {{\mathbb{V}}_{K,B}}, {V'}\in {\mathbb{V}'_{K,B}},
\label{eq:FanoFano}
\end{align}
where (\ref{eq:FanoFano}) follows by (\ref{eq:constraint2}) and Fano's inequality.
Moreover,
\begin{align}
H\left( U^n_{V,B},{{S}_{V}}|U_{\left[ 1:K \right],\left[ 1:B-1 \right]}^{n},{{S}_{V'}} \right)&\geq H\left( U^n_{V,B}|U_{\left[ 1:K \right],\left[ 1:B-1 \right]}^{n},{{S}_{V'}} \right)+H\left( {{S}_{V}}|U_{\left[ 1:K \right],\left[ 1:B \right]}^{n},{{S}_{V'}} \right)\nonumber\\
&\geq H\left( U^n_{V,B}|U^n_{V',B} \right)+H\left( {{S}_{V}}|U_{\left[ 1:K \right],\left[ 1:B \right]}^{n},{{S}_{V'}} \right)\label{eq:Markov}\\
&=H\left( U_{V,B}|U_{V',B} \right)+H\left( {{S}_{V}}|U_{\left[ 1:K \right],\left[ 1:B \right]}^{n},{{S}_{V'}} \right),\label{eq:Fanothird}
\end{align}
where (\ref{eq:Markov}) is due to the fact that $U^n_{V,B}\leftrightarrow U^n_{V',B}\leftrightarrow(U_{\left[ 1:K \right],\left[ 1:B \right]}^{n},{{S}_{V'}})$ form a Markov chain. Continuing from (\ref{eq:firstcont}), 
\begin{align}
&\sum\limits_{V\in {{\mathbb{V}}_{K,B-1}}}{\sum\limits_{{V'}\in {\mathbb{V}'_{K,B-1}}\left[ V \right]}{{{c}_{V|{V'},B-1}}H\left( {{S}_{V}}|U_{\left[ 1:K \right],\left[ 1:B-1 \right]}^{n},{{S}_{V'}} \right)}}\nonumber \\
&\geq n\sum\limits_{V\in {{\mathbb{V}}_{K,B}}}\sum\limits_{{V'}\in {\mathbb{V}'_{K,B}}\left[ V \right]}{{c}_{V|{V'},B}}H\left( U_{V,B}|U_{V',B} \right)\nonumber\\
&\quad+\sum\limits_{V\in {{\mathbb{V}}_{K,B}}}\sum\limits_{{V'}\in {\mathbb{V}'_{K,B}}\left[ V \right]}{{c}_{V|{V'},B}}H\left( {{S}_{V}}|U_{\left[ 1:K \right],\left[ 1:B \right]}^{n},{{S}_{V'}} \right)\nonumber\\
&\quad-n\delta_{\epsilon}\sum\limits_{V\in {{\mathbb{V}}_{K,B}}}\sum\limits_{{V'}\in {\mathbb{V}'_{K,B}}\left[ V \right]}{{c}_{V|{V'},B}}\label{eq:uselot}\\
&\geq n\sum\limits_{V\in {{\mathbb{V}}_{K,B}}}\sum\limits_{{V'}\in {\mathbb{V}'_{K,B}}\left[ V \right]}{{c}_{V|{V'},B}}H\left( U_{V,B}|U_{V',B} \right)\nonumber\\
&\quad+\sum\limits_{V\in {{\mathbb{V}}_{K,B}}}\sum\limits_{{V'}\in {\mathbb{V}'_{K,B}}\left[ V \right]}{{c}_{V|{V'},B}}H\left( {{S}_{V}}|U_{\left[ 1:K \right],\left[ 1:B \right]}^{n},{{S}_{V'}} \right)\nonumber\\
&\quad-n\delta_{\epsilon}\sum\limits_{k=1}^K\sum\limits_{V\in {{\mathbb{V}}_{K,B}}:k\in V}\sum\limits_{{V'}\in {\mathbb{V}'_{K,B}}\left[ V \right]}{{c}_{V|{V'},B}}\nonumber\\
&=n{{f}^{\mathbf{w}}_{{{B}}}}+\sum\limits_{V\in {{\mathbb{V}}_{K,B}}}\sum\limits_{{V'}\in {\mathbb{V}'_{K,B}}\left[ V \right]}{{c}_{V|{V'},B}}H\left( {{S}_{V}}|U_{\left[ 1:K \right],\left[ 1:B \right]}^{n},{{S}_{V'}} \right)-n\delta_{\epsilon}\sum\limits_{k=1}^Kw_k,\label{eq:twoprop}
\end{align}
where (\ref{eq:uselot}) is due to (\ref{eq:FanoFano}) and (\ref{eq:Fanothird}), and (\ref{eq:twoprop}) is due to (\ref{eq:opt}) and (\ref{eq:connection}). Combining (\ref{eq:twoprop}) and the induction hypothesis proves (\ref{eq:induction}) for $\beta=B$.
\end{IEEEproof}

\section{Proof of Theorem \ref{th:main3}}\label{sec:main3}

Theorem~\ref{th:main3} is trivially true when $K=1$. The case $K=2$ is a simple consequence of Lemma \ref{lem:ineq2} and Lemma \ref{lem:outer}. Therefore, only the case $K=3$ remains to be proved.

To this end, we shall give a detailed analysis of ${\mathsf{LP}^{\mathbf{w}}_{3,2}}$. First consider the following related linear program
\begin{align*}
\widetilde{\mathsf{LP}}^{\mathbf{w}}_{3,2}: \max  \quad & \sum\limits_{k=1}^{3}{{{w}_{k}}{{r}_{k,2}}} \\
\operatorname{over} \quad & {{r}_{k,2}},\quad k\in \left[ 1:3 \right], \\
\operatorname{s.t.}  \quad & {{r}_{k,2}}\ge {{\psi }_{\{k\}}},\quad k\in \left[ 1:3 \right] \\
& {{r}_{i,2}}+{{r}_{j,2}}\ge {{\psi }_{\{i\}}}+{{\psi }_{\{i,j\}}}+{{\psi }_{\{j\}}},\quad i,j\in \left[ 1:3 \right], i\neq j,
\end{align*}
where ${\psi }_{V}$, $V\in{{\mathbb{V}}_{3,2}}$, are non-negative real numbers.
We say $\left( {{{\tilde{c}}}_{V,2}},V\in {{\mathbb{V}}_{3,2}}
\right)$ is an optimal Lagrange multiplier of $\widetilde{\mathsf{LP}}^{\mathbf{w}}_{3,2}$ if
\begin{align*}
& \sum\limits_{k=1}^{3}{{{{\tilde{c}}}_{\left\{ k \right\},2}}{{\psi }_{\{k\}}}}+\sum\limits_{i,j\in \left[ 1:3
\right], i<j}{{{{\tilde{c}}}_{\left\{ i,j \right\},2}}\left(
{{\psi }_{\{i\}}}+{{\psi }_{\{i,j\}}}+{{\psi }_{\{j\}}} \right)}={{{\tilde{f}}}^{\mathbf{w}}_{\boldsymbol{\psi
}}}, \\
& \sum\limits_{V\in {{\mathbb{V}}_{3,2}}:k\in V}{{{{\tilde{c}}}_{V,2}}}={{w}_{k}},\quad k\in \left[ 1:3 \right], \\
& {{{\tilde{c}}}_{V,2}}\ge 0,\quad V\in {{\mathbb{V}}_{3,2}},
\end{align*}
where ${{\tilde{f}}^{\mathbf{w}}_{\boldsymbol{\psi }}}$ denotes the optimal value of  $\widetilde{\mathsf{LP}}^{\mathbf{w}}_{3,2}$. One can solve $\widetilde{\mathsf{LP}}^{\mathbf{w}}_{3,2}$ with $\mathbf{w}\in{\mathbb{\mathbb{W} }_{3}}$ by considering 5 different cases (see Table~\ref{tab:lp32new}).

\begin{table*}[!htbp]
	\caption{Linear Program $\widetilde{\mathsf{LP}}^{\mathbf{w}}_{3,2}$}
	\label{tab:lp32new}
	\centering
\begin{tabular}{|c|c|c|c|}
  \hline
  Case & Condition & Optimal Solution & Optimal Lagrange Multiplier  \\
  \hline
1
&
\tabincell{c}{
$ \mathbf{w}\in{{\mathbb{W}}^{\left( 0 \right)}_{3,2 }} $\\
$ {{\psi }_{\{1,2\}}}\le {{\psi }_{\{1,3\}}}+{{\psi }_{\{2,3\}}} $\\
$ {{\psi }_{\{1,3\}}}\le {{\psi }_{\{1,2\}}}+{{\psi }_{\{2,3\}}} $\\
$ {{\psi }_{\{2,3\}}}\le {{\psi }_{\{1,2\}}}+{{\psi }_{\{1,3\}}} $
}
&
\tabincell{c}{
$ \tilde{r}_{1,2}^{\operatorname{opt}}\triangleq{{\psi }_{\{1\}}}+\frac{1}{2}\left( {{\psi }_{\{1,2\}}}+{{\psi }_{\{1,3\}}}-{{\psi }_{\{2,3\}}} \right) $\\
$ \tilde{r}_{2,2}^{\operatorname{opt}}\triangleq{{\psi }_{\{2\}}}+\frac{1}{2}\left( {{\psi }_{\{1,2\}}}-{{\psi }_{\{1,3\}}}+{{\psi }_{\{2,3\}}} \right) $\\
$ \tilde{r}_{3,2}^{\operatorname{opt}}\triangleq{{\psi }_{\{3\}}}+\frac{1}{2}\left( -{{\psi }_{\{1,2\}}}+{{\psi }_{\{1,3\}}}+{{\psi }_{\{2,3\}}} \right)$
}
&
\tabincell{c}{
$ {\tilde{c}_{\left\{ 1,2 \right\},2}}\triangleq\frac{1}{2}\left( {{w}_{1}}+{{w}_{2}}-{{w}_{3}} \right) $\\
$ {\tilde{c}_{\left\{ 1,3 \right\},2}}\triangleq\frac{1}{2}\left( {{w}_{1}}-{{w}_{2}}+{{w}_{3}} \right) $\\
$ {\tilde{c}_{\left\{ 2,3 \right\},2}}\triangleq\frac{1}{2}\left( -{{w}_{1}}+{{w}_{2}}+{{w}_{3}} \right)$\\
$\tilde{c}_{V,2}\triangleq0$, otherwise
}
\\
\hline
2
&
\tabincell{c}{
  $\mathbf{w}\in{{\mathbb{W}}^{\left( 0 \right)}_{3,2 }}$ \\
 $ {{\psi }_{\{1,2\}}}>{{\psi }_{\{1,3\}}}+{{\psi }_{\{2,3\}}} $
}
&
\tabincell{c}{
 $ \tilde{r}_{1,2}^{\operatorname{opt}}\triangleq{{\psi }_{\{1\}}}+{{\psi }_{\{1,3\}}} $ \\
 $ \tilde{r}_{2,2}^{\operatorname{opt}}\triangleq{{\psi }_{\{2\}}}+{{\psi }_{\{1,2\}}}-{{\psi }_{\{1,3\}}} $ \\
 $ \tilde{r}_{3,2}^{\operatorname{opt}}\triangleq{{\psi }_{\{3\}}} $
}
&
\tabincell{c}{
  $ {\tilde{c}_{\left\{ 3 \right\},2}}\triangleq-{{w}_{1}}+{{w}_{2}}+{{w}_{3}} $ \\
 $ {\tilde{c}_{\left\{ 1,2 \right\},2}}\triangleq{{w}_{2}} $ \\
 $ {\tilde{c}_{\left\{ 1,3 \right\},2}}\triangleq{{w}_{1}}-{{w}_{2}} $\\
 $\tilde{c}_{V,2}\triangleq0$, otherwise
} \\
\hline 3 &
\tabincell{c}{
  $ \mathbf{w}\in{{\mathbb{W}}^{\left( 0 \right)}_{3,2 }}$ \\
 $ {{\psi }_{\{1,3\}}}>{{\psi }_{\{1,2\}}}+{{\psi }_{\{2,3\}}} $
}
&
\tabincell{c}{
$ \tilde{r}_{1,2}^{\operatorname{opt}}\triangleq {{\psi }_{\{1\}}}+{{\psi }_{\{1,2\}}} $ \\
$ \tilde{r}_{2,2}^{\operatorname{opt}}\triangleq {{\psi }_{\{2\}}} $ \\
$ \tilde{r}_{3,2}^{\operatorname{opt}}\triangleq {{\psi }_{\{3\}}}+{{\psi }_{\{1,3\}}}-{{\psi }_{\{1,2\}}} $
}
&
\tabincell{c}{
  $ {\tilde{c}_{\left\{ 2 \right\},2}}\triangleq-{{w}_{1}}+{{w}_{2}}+{{w}_{3}} $ \\
 $ {\tilde{c}_{\left\{ 1,2 \right\},2}}\triangleq{{w}_{1}}-{{w}_{3}} $ \\
 $ {\tilde{c}_{\left\{ 1,3 \right\},2}}\triangleq{{w}_{3}} $\\
 $\tilde{c}_{V,2}\triangleq0$, otherwise
} \\
\hline 4 &
\tabincell{c}{
  $ \mathbf{w}\in{{\mathbb{W}}_3}$ \\
${{\psi }_{\{2,3\}}}>{{\psi }_{\{1,2\}}}+{{\psi }_{\{1,3\}}}$}
&
\tabincell{c}{
$ \tilde{r}_{1,2}^{\operatorname{opt}}\triangleq {{\psi }_{\{1\}}} $ \\
$ \tilde{r}_{2,2}^{\operatorname{opt}}\triangleq {{\psi }_{\{2\}}}+{{\psi }_{\{1,2\}}} $ \\
$ \tilde{r}_{3,2}^{\operatorname{opt}}\triangleq {{\psi }_{\{3\}}}+{{\psi }_{\{2,3\}}}-{{\psi }_{\{1,2\}}} $
}
&
\tabincell{c}{
  $ {\tilde{c}_{\left\{ 1 \right\},2}}\triangleq{{w}_{1}}-{{w}_{2}}+{{w}_{3}} $ \\
 $ {\tilde{c}_{\left\{ 1,2 \right\},2}}\triangleq{{w}_{2}}-{{w}_{3}} $ \\
 $ {\tilde{c}_{\left\{ 2,3 \right\},2}}\triangleq{{w}_{3}} $\\
 $\tilde{c}_{V,2}\triangleq0$, otherwise
} \\
\hline 5 &
\tabincell{c}{
  $ \mathbf{w}\in{{\mathbb{W}}^{\left( 1 \right)}_{3,2 }}$ \\
 $ {{\psi }_{\{2,3\}}}\le {{\psi }_{\{1,2\}}}+{{\psi }_{\{1,3\}}} $ \\
}
&
\tabincell{c}{
$ \tilde{r}_{1,2}^{\operatorname{opt}}\triangleq {{\psi }_{\{1\}}} $ \\
$ \tilde{r}_{2,2}^{\operatorname{opt}}\triangleq  {{\psi }_{\{2\}}}+{{\psi }_{\{1,2\}}} $ \\
$ \tilde{r}_{3,2}^{\operatorname{opt}}\triangleq  {{\psi }_{\{3\}}}+{{\psi }_{\{1,3\}}} $
}
&
\tabincell{c}{
  $ {\tilde{c}_{\left\{ 1 \right\},2}}\triangleq{{w}_{1}}-{{w}_{2}}-{{w}_{3}} $ \\
 $ {\tilde{c}_{\left\{ 1,2 \right\},2}}\triangleq{{w}_{2}} $ \\
 $ {\tilde{c}_{\left\{ 1,3 \right\},2}}\triangleq{{w}_{3}} $\\
 $\tilde{c}_{V,2}\triangleq0$, otherwise
} \\
  \hline
\end{tabular}
\end{table*}

Now set
\begin{align*}
&{{\psi }_{\{k\}}}
\triangleq H\left( {{U}_{k,2}}|{{U}_{\nu_k,2}}\right), \quad k\in \left[ 1:3 \right],
\\
&{{\psi }_{\{i,j\}}}
\triangleq
\max \left\{ H\left( {{U}_{i,2}},{{U}_{j,2}} \right)-{{\psi }_{i}}-{{\psi }_{j}},0 \right\}, \quad i,j\in \left[ 1:3 \right], i \neq j,
\end{align*}
where ${{\nu }_{k}}$ is a maximizer of $\max_{\nu \in \left[ 1:3 \right]\backslash\{k\}}H\left( {{U}_{k,2}}|{{U}_{\nu ,2}} \right)$. Moreover, define
\begin{align*}
{\tilde{{\mathcal{R}}}_{3,2}}   & \triangleq\left\{ \left( {{r}_{k,2}},k\in \left[ 1:3 \right] \right): {{r}_{k,2}}\ge {{\psi }_{\{k\}}}, k\in\left[1:3\right],\right.\\
&\left.\hspace{1.42in} {{r}_{i,2}}+{{r}_{j,2}}\ge {{\psi }_{\{i\}}}+{{\psi }_{\{i,j\}}}+{{\psi }_{\{j\}}},   i,j\in \left[ 1:3 \right], i\ne j \right\}.
\end{align*}


One can prove via direct verification that ${{\mathcal{R}}_{3,2}}$ coincides with ${\tilde{{\mathcal{R}}}_{3,2}}$.

\begin{lemma}
\label{lem:lp32_region}
${{\mathcal{R}}_{3,2}}={\tilde{{\mathcal{R}}}_{3,2}}$.
\end{lemma}

\begin{figure}[thbp!]
	\centering
		\includegraphics{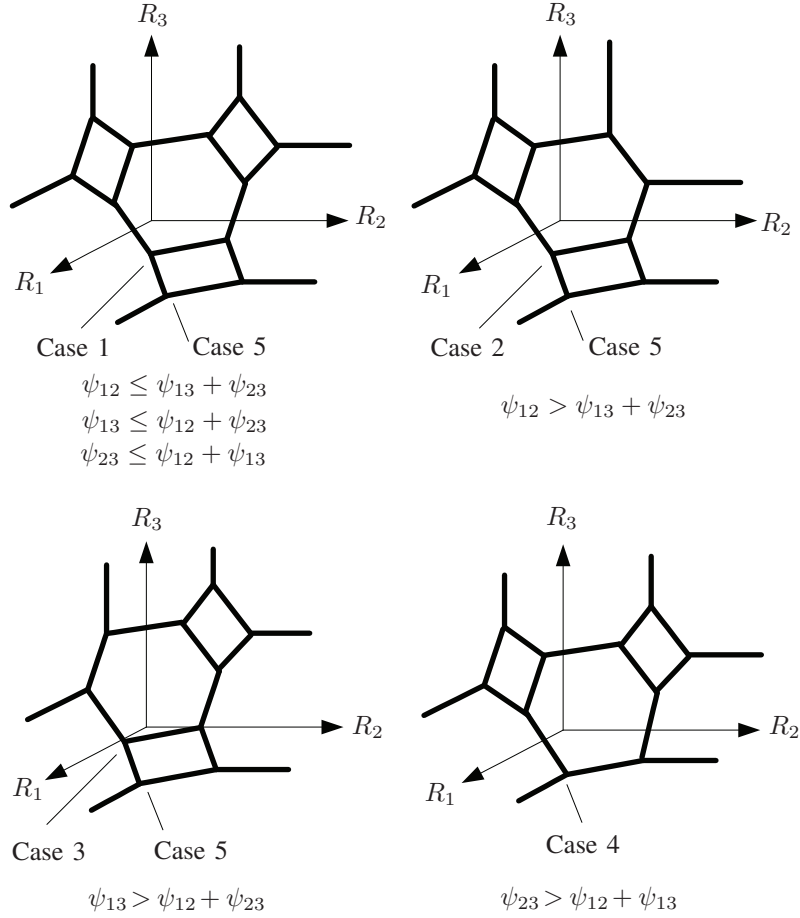}
	\caption{Illustrations of ${{\mathcal{R}}_{3,2}}$ (i.e., ${\tilde{{\mathcal{R}}}_{3,2}}$) and optimal solutions in Table~\ref{tab:lp32new}.}
	\label{fig:rateregion3}
\end{figure}

See Fig.~\ref{fig:rateregion3} for illustrations of ${{\mathcal{R}}_{3,2}}$ (i.e., ${\tilde{{\mathcal{R}}}_{3,2}}$), where the optimal solutions in Table~\ref{tab:lp32new} are highlighted.

\begin{lemma}
\label{lem:psi}
Let $i,j,k$ be three distinct integers in $\left[ 1:3 \right]$.

(1) ${{\psi }_{\{i\}}}+{{\psi }_{\{i,j\}}}+{{\psi }_{\{j\}}}=\begin{cases}
   H\left( {{U}_{i,2}},{{U}_{j,2}} \right), & H\left( {{U}_{i,2}},{{U}_{j,2}} \right)\ge {{\psi }_{\{i\}}}+{{\psi }_{\{j\}}},  \\
   H\left( {{U}_{i,2}}|{{U}_{k,2}} \right)+H\left( {{U}_{j,2}}|{{U}_{k,2}} \right), & H\left( {{U}_{i,2}},{{U}_{j,2}} \right)<{{\psi }_{\{i\}}}+{{\psi }_{\{j\}}}.
\end{cases}$

(2) If ${{\psi }_{\{i,j\}}}>{{\psi }_{\{i,k\}}}+{{\psi }_{\{j,k\}}}$, then ${{\psi }_{\{i\}}}+{{\psi }_{\{i,j\}}}+{{\psi }_{\{j\}}}=H\left( {{U}_{i,2}},{{U}_{j,2}} \right)$.

(3) If ${{\psi }_{\{i\}}}+{{\psi }_{\{i,j\}}}+{{\psi }_{\{j\}}}=H\left( {{U}_{i,2}}|{{U}_{k,2}} \right)+H\left( {{U}_{j,2}}|{{U}_{k,2}} \right)$, then
\begin{align*}
&{{\psi }_{\{i\}}}+{{\psi }_{\{i,k\}}}+{{\psi }_{\{k\}}} =H\left( {{U}_{i,2}},{{U}_{k,2}} \right) , \\
&{{\psi }_{\{j\}}}+{{\psi }_{\{j,k\}}}+{{\psi }_{\{k\}}}=H\left( {{U}_{j,2}},{{U}_{k,2}} \right).
\end{align*}
\end{lemma}
\begin{IEEEproof}
See Appendix \ref{app:psi}.
\end{IEEEproof}

According to Lemma~\ref{lem:psi}, we have the following four cases for ${{\psi }_{\{i\}}}+{{\psi }_{\{i,j\}}}+{{\psi }_{\{j\}}}$ ($i,j\in \left[ 1:3 \right],i\ne j$):
\begin{enumerate}[\quad (C{a}se A)]
\item
\begin{align*}
&{{\psi }_{\{1\}}}+{{\psi }_{\{1,2\}}}+{{\psi }_{\{2\}}}=H\left( {{U}_{1,2}},{{U}_{2,2}} \right), \\
&{{\psi }_{\{1\}}}+{{\psi }_{\{1,3\}}}+{{\psi }_{\{3\}}}=H\left( {{U}_{1,2}},{{U}_{3,2}} \right), \\
&{{\psi }_{\{2\}}}+{{\psi }_{\{2,3\}}}+{{\psi }_{\{3\}}}=H\left( {{U}_{2,2}},{{U}_{3,2}} \right);
\end{align*}
\item
\begin{align*}
&{{\psi }_{\{1\}}}+{{\psi }_{\{1,2\}}}+{{\psi }_{\{2\}}}=H\left( {{U}_{1,2}},{{U}_{2,2}} \right), \\
&{{\psi }_{\{1\}}}+{{\psi }_{\{1,3\}}}+{{\psi }_{\{3\}}}=H\left( {{U}_{1,2}},{{U}_{3,2}} \right), \\
&{{\psi }_{\{2\}}}+{{\psi }_{\{2,3\}}}+{{\psi }_{\{3\}}}=H\left( {{U}_{2,2}}|{{U}_{1,2}} \right)+H\left( {{U}_{3,2}}|{{U}_{1,2}} \right)>H\left( {{U}_{2,2}},{{U}_{3,2}} \right);
\end{align*}
\item
\begin{align*}
&{{\psi }_{\{1\}}}+{{\psi }_{\{1,2\}}}+{{\psi }_{\{2\}}}=H\left( {{U}_{1,2}},{{U}_{2,2}} \right), \\
&{{\psi }_{\{1\}}}+{{\psi }_{\{1,3\}}}+{{\psi }_{\{3\}}}=H\left( {{U}_{1,2}}|{{U}_{2,2}} \right)+H\left( {{U}_{3,2}}|{{U}_{2,2}} \right)>H\left( {{U}_{1,2}},{{U}_{3,2}} \right), \\
&{{\psi }_{\{2\}}}+{{\psi }_{\{2,3\}}}+{{\psi }_{\{3\}}}=H\left( {{U}_{2,2}},{{U}_{3,2}} \right);
\end{align*}
\item
\begin{align*}
&{{\psi }_{\{1\}}}+{{\psi }_{\{1,2\}}}+{{\psi }_{\{2\}}}=H\left( {{U}_{1,2}}|{{U}_{3,2}} \right)+H\left( {{U}_{2,2}}|{{U}_{3,2}} \right)>H\left( {{U}_{1,2}},{{U}_{2,2}} \right), \\
&{{\psi }_{\{1\}}}+{{\psi }_{\{1,3\}}}+{{\psi }_{\{3\}}}=H\left( {{U}_{1,2}},{{U}_{3,2}} \right), \\
&{{\psi }_{\{2\}}}+{{\psi }_{\{2,3\}}}+{{\psi }_{\{3\}}}=H\left( {{U}_{2,2}},{{U}_{3,2}} \right).
\end{align*}
\end{enumerate}

Now one can readily solve ${\mathsf{LP}^{\mathbf{w}}_{3,2}}$ with $\mathbf{w}\in {{\mathbb{W}}_{3}}$ by considering all possible combinations of these four cases and those in Table~\ref{tab:lp32new} (i.e., Cases 1--5). For example, consider the scenario where Case 2 and Case C are simultaneously satisfied (henceforth called Case 2C). It can be verified that
\begin{align*}
&{{\psi }_{\{1\}}}=H\left(U_{1,2}|U_{2,2}\right),\\
&{{\psi }_{\{3\}}}=H\left(U_{3,2}|U_{2,2}\right),\\
&{{\psi }_{\{1,3\}}}=0,\\
&{{\psi }_{\{1\}}}+{{\psi }_{\{1,2\}}}+{{\psi }_{\{2\}}}=H\left( {{U}_{1,2}},{{U}_{2,2}} \right).
\end{align*}
For $\left( \tilde{r}_{k,2}^{\operatorname{opt}},k\in \left[ 1:3 \right] \right)$ in Table~\ref{tab:lp32new}, we have
\begin{align*}
\tilde{r}_{1,2}^{\operatorname{opt}}& = {{\psi }_{\{1\}}}+{{\psi }_{\{1,3\}}}\\
&=H\left( {{U}_{1,2}}|{{U}_{2,2}} \right), \\
\tilde{r}_{2,2}^{\operatorname{opt}} &={{\psi }_{\{2\}}}+{{\psi }_{\{1,2\}}}-{{\psi }_{\{1,3\}}} \\
&= \left( {{\psi }_{\{1\}}}+{{\psi }_{\{1,2\}}}+{{\psi }_{\{2\}}} \right)-\left( {{\psi }_{\{1\}}}+{{\psi }_{\{1,3\}}} \right)\\
&=H\left( {{U}_{2,2}} \right), \\
\tilde{r}_{3,2}^{\operatorname{opt}} &= {{\psi }_{\{3\}}}\\
&=H\left( {{U}_{3,2}}|{{U}_{2,2}} \right).
\end{align*}
In view of  Lemma \ref{lem:lp32_region}, $\left( r_{k,2}^{\operatorname{opt}},k\in \left[ 1:K \right] \right)$ with $r_{k,2}^{\operatorname{opt}}\triangleq\tilde{r}_{k,2}^{\operatorname{opt}}$, $k\in \left[ 1:K \right]$, is an optimal solution of  ${\mathsf{LP}^{\mathbf{w}}_{3,2}}$. Therefore, the optimal value of ${\mathsf{LP}^{\mathbf{w}}_{3,2}}$ is given by
\begin{align*}
{{f}^{\mathbf{w}}_{{{2}}}} &={{w}_{1}}H\left( {{U}_{1,2}}|{{U}_{2,2}} \right)+{{w}_{2}}H\left( {{U}_{2,2}} \right)+{{w}_{3}}H\left( {{U}_{3,2}}|{{U}_{2,2}} \right).
\end{align*}
Moreover, $\left( {{c}_{V|{V'},2 }},V\in {{\mathbb{V}}_{3,2 }},{V'}\in {\mathbb{V}'_{3,2}}\left[ V \right] \right)$ with
\begin{align*}
&c_{\{1\}|\{2\},2}\triangleq w_1-w_2,\\
&c_{\{3\}|\{2\},2}\triangleq w_3,\\
&c_{\{1,2\}|\emptyset,2}\triangleq w_2,\\
&{{c}_{V|{V'},2 }}\triangleq0,\quad\mbox{otherwise},
\end{align*}
is an optimal Lagrange multiplier of ${\mathsf{LP}^{\mathbf{w}}_{3,2}}$.

\begin{table*}[!htbp]
        \addtolength{\tabcolsep}{-3pt}
        \caption{Optimal Lagrange Multipliers of $\mathsf{LP}_{3,2}$ for All Possible Cases}
        \label{tab:lp32lag}
        \centering
\begin{tabular}{|c|c|c|c|c|c|c|}
  \hline
Case & ${{c}_{\left\{ 1 \right\}|\left\{ {{\nu }_{1}} \right\},2}}$ & ${{c}_{\left\{ 2 \right\}|\left\{ {{\nu}_{2}} \right\},2}}$ & ${{c}_{\left\{ 3 \right\}|\left\{ {{\nu }_{3}} \right\},2}}$ & ${{c}_{\left\{ 1,2 \right\}|\emptyset,2}}$ & ${{c}_{\left\{ 1,3 \right\}|\emptyset,2}}$ & ${{c}_{\left\{ 2,3 \right\}|\emptyset,2}}$ \\ \hline
1A & $0$ & $0$ & $0$ & $\frac{1}{2}\left( {{w}_{1}}+{{w}_{2}}-{{w}_{3}} \right)$ & $\frac{1}{2}\left( {{w}_{1}}-{{w}_{2}}+{{w}_{3}} \right)$ & $\frac{1}{2}\left( -{{w}_{1}}+{{w}_{2}}+{{w}_{3}} \right)$ \\ \hline
1B & $0$ & $\frac{1}{2}\left( -{{w}_{1}}+{{w}_{2}}+{{w}_{3}} \right)$ & $\frac{1}{2}\left( -{{w}_{1}}+{{w}_{2}}+{{w}_{3}} \right)$ & $\frac{1}{2}\left( {{w}_{1}}+{{w}_{2}}-{{w}_{3}} \right)$ & $\frac{1}{2}\left( {{w}_{1}}-{{w}_{2}}+{{w}_{3}} \right)$ & $0$ \\ \hline
1C & $\frac{1}{2}\left( {{w}_{1}}-{{w}_{2}}+{{w}_{3}} \right)$ & $0$ & $\frac{1}{2}\left( {{w}_{1}}-{{w}_{2}}+{{w}_{3}} \right)$ & $\frac{1}{2}\left( {{w}_{1}}+{{w}_{2}}-{{w}_{3}} \right)$ & $0$ & $\frac{1}{2}\left( -{{w}_{1}}+{{w}_{2}}+{{w}_{3}} \right)$ \\ \hline
1D & $\frac{1}{2}\left( {{w}_{1}}+{{w}_{2}}-{{w}_{3}} \right)$ & $\frac{1}{2}\left( {{w}_{1}}+{{w}_{2}}-{{w}_{3}} \right)$ & $0$ & $0$ & $\frac{1}{2}\left( {{w}_{1}}-{{w}_{2}}+{{w}_{3}} \right)$ & $\frac{1}{2}\left( -{{w}_{1}}+{{w}_{2}}+{{w}_{3}} \right)$ \\ \hline
2A & $0$ & $0$ & $ -{{w}_{1}}+{{w}_{2}}+{{w}_{3}}$ & ${{w}_{2}}$ & ${{w}_{1}}-{{w}_{2}}$ & $0$ \\ \hline
2B & $0$ & $0$ & $ -{{w}_{1}}+{{w}_{2}}+{{w}_{3}}$ & ${{w}_{2}}$ & ${{w}_{1}}-{{w}_{2}}$ & $0$ \\ \hline
2C & ${{w}_{1}}-{{w}_{2}}$ & $0$ & ${{w}_{3}}$ & ${{w}_{2}}$ & $0$ & $0$ \\ \hline
3A & $0$ & $-{{w}_{1}}+{{w}_{2}}+{{w}_{3}}$ & $0$ & ${{w}_{1}}-{{w}_{3}}$ & ${{w}_{3}}$ & $0$ \\ \hline
3B & $0$ & $-{{w}_{1}}+{{w}_{2}}+{{w}_{3}}$ & $0$ & ${{w}_{1}}-{{w}_{3}}$ & ${{w}_{3}}$ & $0$ \\ \hline
3D & ${{w}_{1}}-{{w}_{3}}$ & ${{w}_{2}}$ & $0$ & $0$ & ${{w}_{3}}$ & $0$ \\ \hline
4A & ${{w}_{1}}-{{w}_{2}}+{{w}_{3}}$ & $0$ & $0$ & ${{w}_{2}}-{{w}_{3}}$ & $0$ & ${{w}_{3}}$ \\ \hline
4C & ${{w}_{1}}-{{w}_{2}}+{{w}_{3}}$ & $0$ & $0$ & ${{w}_{2}}-{{w}_{3}}$ & $0$ & ${{w}_{3}}$ \\ \hline
4D & ${{w}_{1}}$ & ${{w}_{2}}-{{w}_{3}}$ & $0$ & $0$ &$0$ & ${{w}_{3}}$ \\ \hline
5A & ${{w}_{1}}-{{w}_{2}}-{{w}_{3}}$ & $0$ & $0$ & ${{w}_{2}}$ & ${{w}_{3}}$ & $0$ \\ \hline
5B & ${{w}_{1}}-{{w}_{2}}$ & $0$ & ${{w}_{3}}$ & ${{w}_{2}}$ & $0$ & $0$ \\ \hline
5C & ${{w}_{1}}-{{w}_{2}}-{{w}_{3}}$ & $0$ & $0$ & ${{w}_{2}}$ & ${{w}_{3}}$ & $0$ \\ \hline
5D & ${{w}_{1}}-{{w}_{3}}$ & ${{w}_{2}}$ & $0$ & $0$ & ${{w}_{3}}$ & $0$ \\ \hline
\end{tabular}
\end{table*}




One can obtain the following lemma by analyzing the other combinations in the same manner. It is worth mentioning that not all combinations are possible. Specifically,  Cases 2D, 3C, and  4B violate Lemma~\ref{lem:psi}(2), so such combinations are void. 

\begin{lemma}
\label{lem:lp32}
For linear program $\mathsf{LP}^{\mathbf{w}}_{3,2}$ with $\mathbf{w}\in {{\mathbb{W}}_{3}}$,
$\left( {{c}_{V|{V'},2}}, V\in {{\mathbb{V}}_{3,2}},{V'}\in {\mathbb{V}'_{3,2}}\left[ V \right]  \right)$ in Table~\ref{tab:lp32lag}\footnote{We set $c_{\{k\}|\{k'\},2}=0$ for $k'\neq\nu_k$.} is an optimal Lagrange multiplier.
The general case $\mathbf{w}\in \mathbb{R}_{+}^{3}$  can be reduced to the case $\mathbf{w}\in{\mathbb{\mathbb{W} }_{3}}$ via suitable relabelling. 
\end{lemma}


The next result shows that (\ref{eq:star}) holds when $K=3$, which, together with (\ref{eq:innerbound}) and Lemma \ref{lem:outer}, completes the proof of Theorem \ref{th:main3}.

\begin{lemma}
\label{lem:ineq3}
Let $\left( {{c}_{V,1}}, V\in {{\mathbb{V}}_{3,1}} \right)$, $\left( {{c}_{V|{V'},2}},V\in {{\mathbb{V}}_{3,2}},{V'}\in {\mathbb{V}'_{3,2}}\left[ V \right] \right)$, and $\left( {{c}_{V,3}}, V\in {{\mathbb{V}}_{3,3}} \right)$ be the optimal multipliers in Lemma~\ref{lem:lpk1}, Lemma~\ref{lem:lpkk}, and Lemma~\ref{lem:lp32}, respectively. We have
\begin{align}
\sum\limits_{V\in {{\mathbb{V}}_{3,1}}}{{{c}_{V,1}}H\left( {{X}_{V}} \right)}&\ge \sum\limits_{V\in {{\mathbb{V}}_{3,2}}}{\sum\limits_{{V'}\in {\mathbb{V}'_{3,2}}\left[ V \right]}{{{c}_{V|{V'},2}}H\left( {{X}_{V}}|{{X}_{V'}} \right)}}\label{eq:leftinequality}\\
&\ge \sum\limits_{V\in {{\mathbb{V}}_{3,3}}}{{{c}_{V,3}}H\left( {{X}_{V}}|{{X}_{\left[ 1:3 \right]\backslash V}} \right)}\label{eq:rightinequality}
\end{align}
for all $X_{\left[1:3\right]}$.
\end{lemma}
\begin{IEEEproof}
Note that (\ref{eq:leftinequality}) follows from Lemma \ref{lem:ineq2}. The proof of (\ref{eq:rightinequality}) is relegated to  Appendix~\ref{app:rightinequality}.
\end{IEEEproof}

\section{Proof of Theorem~\ref{th:main_sym}}
\label{sec:prove_sym}

The proof of Theorem~\ref{th:main_sym} also largely follows the general approach outlined in Section~\ref{sec:prove}. However, due to the symmetry assumption, some simplifications are possible.


\subsection{Linear Program}
\label{sec:r_sym}

When the distribution of ${{U}_{\left[ 1:K \right],\left[ 1:K \right]}}$ is symmetrical entropy-wise, $H\left( {{U}_{V,\alpha }}|{{U}_{{V'},\alpha }} \right)$ depends on $V\in {{\mathbb{V}}_{K,\alpha }}$ and $V'\in {\mathbb{V}'_{K,\alpha }}\left[ V \right]$ only through $|V|$; for this reason, we shall denote it as ${{H}_{\left| V \right|,\alpha }}$ and rewrite
 ${\mathsf{LP}^{\mathbf{w}}_{K,\alpha}}$ in the following simpler form
\begin{align*}
{\overline{\mathsf{LP}}^{\mathbf{w}}_{K,\alpha }}: \min \quad &
\sum\limits_{k=1}^{K}{{{w}_{k}}{{r}_{k,\alpha }}} \\
\operatorname{over} \quad &
{{r}_{k,\alpha }}, \quad k\in \left[ 1:K \right], \\
\operatorname{s.t.} \quad &
\sum\limits_{k\in V}{{{r}_{k,\alpha }}}\ge {{H}_{\left| V \right|,\alpha }},\quad V\in {{\mathbb{V}}_{K,\alpha }}.
\end{align*}

\begin{definition}
\label{def:lagrangesymmetric}
 We say $\left( {{c}_{V,\alpha }},V\in {{\mathbb{V}}_{K,\alpha }} \right)$ is an optimal Lagrange multiplier of ${\overline{\mathsf{LP}}^{\mathbf{w}}_{K,\alpha }}$ with $\mathbf{w}\in \mathbb{R}_{+}^{K}$ if 
\begin{align}
&\sum\limits_{V\in {{\mathbb{V}}_{K,\alpha }}}{{c}_{V,\alpha }}{{H}_{\left| V \right|,\alpha }}= {\overline{{f}}^{\mathbf{w}}_{{{\alpha }}}}, \label{eq:lag_sym_f}
\\
&\sum\limits_{V\in {{\mathbb{V}}_{K,\alpha }}:k\in V}{{{c}_{V,\alpha }}} = {{w}_{k}},\quad k\in \left[ 1:K \right],\label{eq:lag_sym_w}
\\
&{{c}_{V,\alpha }} \ge 0,\quad V\in {{\mathbb{V}}_{K,\alpha }}, \label{eq:lag_sym_p}
\end{align}
where ${\overline{{f}}^{\mathbf{w}}_{{{\alpha }}}}$ denotes the optimal value of ${\overline{\mathsf{LP}}^{\mathbf{w}}_{K,\alpha }}$.
\end{definition}




For $l\in\left[0:\alpha-1\right]$, define
\begin{align*}
r^{\left(l\right)}_{k,\alpha }\triangleq\begin{cases}
   {{H}_{k,\alpha }}-{{H}_{k-1,\alpha }}, & k\in\left[1: {{l}}\right],  \\
   \frac{{{H}_{\alpha ,\alpha }}-{{H}_{{{l}},\alpha }}}{\alpha -{{l}}}, & k\in\left[l+1:K\right],
\end{cases}
\end{align*}
where $H_{0,\alpha}\triangleq0$.

\begin{lemma}
\label{lem:opt_in_region}
$\left( {{r}^{\left(l\right)}_{k,\alpha }},k\in \left[ 1:K \right] \right)\in {{\mathcal{R}}_{K,\alpha }}$.
\end{lemma}
\begin{IEEEproof}
See Appendix \ref{app:opt_in_region}.
\end{IEEEproof}


For $\alpha \in \left[ 1:K \right]$ and $l\in \left[ 0:\alpha -1 \right]$, define
\begin{align*}
{{\Omega }^{\left( l \right)}_{K,\alpha }}\triangleq\left\{ V\subseteq \left[ 1:K \right]:\left| V \right|=\alpha ,\left[ 1:l \right]\subseteq V \right\}.
\end{align*}
Recall that ${{\mathbb{W}}^{\left( 0 \right)}_{K,\alpha }},\cdots,{{\mathbb{W}}^{\left( \alpha-1 \right)}_{K,\alpha }}$  form a partition of $\mathbb{W}_{K}$. For $\mathbf{w}\in {{\mathbb{W}}_{K}}$ and $\alpha\in[1:K]$, let $l^{\mathbf{w}}_{\alpha}$ denote the unique integer in $[0:\alpha-1]$ such that $\mathbf{w}\in{{\mathbb{W}}^{\left( l^{\mathbf{w}}_{\alpha} \right)}_{K,\alpha }}$; it is easy to verify that
\begin{align*}
0={{l}^{\mathbf{w}}_{1}}\le {{l}^{\mathbf{w}}_{2}}\le \cdots \le {{l}^{\mathbf{w}}_{K-1}}\le {{l}^{\mathbf{w}}_{K}}.
\end{align*}
Moreover, for  $\mathbf{w}\in {{\mathbb{W}}_{K}}$ and $\alpha\in[1:K]$, define
\begin{align*}
{\lambda^\mathbf{w}_{\alpha }}\triangleq\frac{1}{\alpha -{{l}^\mathbf{w}_{\alpha }}}\sum\limits_{k={{l}^\mathbf{w}_{\alpha }}+1}^{K}{{{w}_{k}}}
\end{align*}
and
\begin{align}
{{\mathbb{C}}^{\mathbf{w}}_{K,\alpha }} \triangleq &\left\{ \left( {{c}_{V,\alpha }}:V\in {{\mathbb{V}}_{K,\alpha }} \right): \right.
& 
\nonumber \\
& {{c}_{\left[ 1:k \right],\alpha }}={{w}_{k}}-{{w}_{k+1}}, \quad
k\in \left[ 1:{{l}^{\mathbf{w}}_{\alpha }}-1 \right],
\label{eq:def_mathbb_c1} \\
& {{c}_{\left[ 1:{{l}^{\mathbf{w}}_{\alpha }} \right],\alpha }}={{w}_{{{l}^{\mathbf{w}}_{\alpha }}}}-{\lambda^\mathbf{w}_{\alpha }},\label{eq:need_verify_1} \\
& {{c}_{V,\alpha }}\ge 0,\quad  V\in \Omega _{K,\alpha }^{\left( {{l}^{\mathbf{w}}_{\alpha }} \right)},
\label{eq:need_verify_2} \\
& {{c}_{V,\alpha }}=0,\quad  \text{otherwise},\label{eq:otherwise} \\
& \sum\limits_{V\in \Omega _{K,\alpha }^{\left( {{l}^{\mathbf{w}}_{\alpha }} \right)}:k\in V}{{{c}_{V,\alpha }}}={{w}_{k}},\quad  k\in \left[ {{l}^{\mathbf{w}}_{\alpha }}+1:K \right] \}.
\label{eq:need_verify_3}
\end{align}
Note that (\ref{eq:def_mathbb_c1}) and (\ref{eq:need_verify_1}) are void when ${{l}^{\mathbf{w}}_{\alpha }}=0$. The definition of ${{\mathbb{C}}^{\mathbf{w}}_{K,\alpha }}$ can be extended to the case $\mathbf{w}\in \mathbb{R}_{+}^{K}$ through suitable labelling.

\begin{lemma}
\label{lem:prepare_lag_sym}
For any $\mathbf{w}\in {{\mathbb{W}}_{K}}$, $\left( {{c}_{V,\alpha-1 }},V\in {{\mathbb{V}}_{K,\alpha-1 }} \right)\in {{\mathbb{C}}^{\mathbf{w}}_{K,\alpha-1 }}$, and $\left( {{c}_{V,\alpha }},V\in {{\mathbb{V}}_{K,\alpha }} \right)\in {{\mathbb{C}}^{\mathbf{w}}_{K,\alpha }}$,
\begin{align}
&\sum\limits_{V\in {{\Omega }^{\left({{l}^{\mathbf{w}}_{\alpha }}\right)}_{K,\alpha }}}{{{c}_{V,\alpha}}}={\lambda^\mathbf{w}_{\alpha }},\label{eq:partial_sum}\\
&\sum\limits_{V\in {{\mathbb{V}}_{K,\alpha }}:k\in V}{{{c}_{V,\alpha }}}={{w}_{k}},\quad k\in \left[ 1:K \right],\label{eq:lag_sym_w_reprise}\\
&{{c}_{V,\alpha }}\ge 0,\quad V\in {{\mathbb{V}}_{K,\alpha }},\label{eq:lag_sym_p_reprise}\\
&{{c}_{\left[ 1:l_{\alpha -1}^{\mathbf{w}}  \right],\alpha }}-{{c}_{\left[ 1:l_{\alpha -1}^{\mathbf{w}}  \right],\alpha -1}}=\theta _{\alpha }^{\mathbf{w}}\ge 0,\label{eq:moved}\\
&\sum\limits_{V\in {{\mathbb{V}}_{K,\alpha }}}{{{c}_{V,\alpha }}}=
\begin{cases}
   \frac{1}{\alpha }\sum\limits_{k=1}^{K}{{{w}_{k}}}, & l^{\mathbf{w}}_\alpha=0,  \\
   {{w}_{1}}, & l^{\mathbf{w}}_\alpha>0.\label{eq:full_sum}
\end{cases}
\end{align}
where
\begin{align}
\theta _{\alpha }^{\mathbf{w}}\triangleq\left( \lambda _{\alpha }^{\mathbf{w}}-\sum\limits_{k=l_{\alpha -1}^{\mathbf{w}}+1}^{l_{\alpha }^{\mathbf{w}}}{\left( \alpha-1 -k \right){{c}_{\left[ 1:k \right],\alpha }}} \right)\frac{1}{\alpha-1 -l_{\alpha -1}^{\mathbf{w}}}.\label{eq:deftheta}
\end{align}
\end{lemma}
\begin{IEEEproof}
See Appendix~\ref{app:prepare_lag_sym}.
\end{IEEEproof}

The main result of Section \ref{sec:r_sym} is as follows.

\begin{lemma}
\label{lem:lp_sym}
For linear program ${\overline{\mathsf{LP}}^{\mathbf{w}}_{K,\alpha }}$ with $\mathbf{w}\in {{\mathbb{W} }
_{K}}$, $\left( {{r}^{\left(l^{\mathbf{w}}_{\alpha}\right)}_{k,\alpha }},k\in \left[ 1:K \right] \right)$
is an optimal solution, and every $\left( {{c}_{V,\alpha }},V\in {{\mathbb{V}}_{K,\alpha }} \right)\in \mathbb{C}^{\mathbf{w}}_{K,\alpha }$ is an optimal Lagrange multiplier.
\end{lemma}
\begin{IEEEproof}
In view of Lemma~\ref{lem:opt_in_region}, we have $\left( r_{k,\alpha }^{\left( l^{\mathbf{w}}_\alpha \right)}, k\in \left[ 1:K \right] \right)\in {{\mathcal{R}}_{K,\alpha }}$. Consider an arbitrary $\left( {{c}_{V,\alpha }},V\in {{\mathbb{V}}_{K,\alpha }} \right)\in \mathbb{C}^{\mathbf{w}}_{K,\alpha }$. It follows from Lemma~\ref{lem:prepare_lag_sym} that  $\left( {{c}_{V,\alpha }},V\in {{\mathbb{V}}_{K,\alpha }} \right)$ satisfies (\ref{eq:lag_sym_w}) and~(\ref{eq:lag_sym_p}).
Note that
\begin{align}
\sum\limits_{k=1}^{K}{{{w}_{k}}r_{k,\alpha }^{\left( l^{\mathbf{w}}_\alpha \right)}}&=\sum\limits_{k=1}^{{{l}^{\mathbf{w}}_{\alpha }}}{{{w}_{k }}\left( {{H}_{k,\alpha }}-{{H}_{k-1,\alpha }} \right)}+\sum\limits_{k={{l}^{\mathbf{w}}_{\alpha }}+1}^{K}{{{w}_{k}}}\frac{{{H}_{\alpha ,\alpha }}-{{H}_{{{l}^{\mathbf{w}}_{\alpha }},\alpha }}}{\alpha -{{l}^{\mathbf{w}}_{\alpha }}} \nonumber\\
&= \sum\limits_{k=1}^{{{l}^{\mathbf{w}}_{\alpha }}-1}{\left( {{w}_{k }}-{{w}_{k+1 }} \right){{H}_{k,\alpha }}}+\left( {{w}_{{{l}^{\mathbf{w}}_{\alpha }} }}-{\lambda^\mathbf{w}_{\alpha }} \right){{H}_{{{l}^{\mathbf{w}}_{\alpha }},\alpha }}+{\lambda^\mathbf{w}_{\alpha }} {{H}_{\alpha ,\alpha }} \nonumber\\
 &= \sum\limits_{k=1}^{{{l}^{\mathbf{w}}_{\alpha }}-1}{{{c}_{\left[ 1:k \right],\alpha }}{{H}_{k,\alpha }}}+{{c}_{\left[ 1:{{l}^{\mathbf{w}}_{\alpha }} \right],\alpha }}{{H}_{{{l}^{\mathbf{w}}_{\alpha }},\alpha }}+\sum\limits_{V\in \Omega _{K,\alpha }^{\left( {{l}^{\mathbf{w}}_{\alpha }} \right)}}{{{c}_{V,\alpha }}}{{H}_{\alpha ,\alpha }} \label{eq:use_in_cl}\\
 &= \sum\limits_{V\in {{\mathbb{V}}_{K,\alpha }}}{{{c}_{V,\alpha }}{{H}_{\left| V \right|,\alpha }}}, \nonumber
\end{align}
where (\ref{eq:use_in_cl}) is due to (\ref{eq:partial_sum}). 
On the other hand, for any $\left( {{r}_{k,\alpha }}:k\in \left[ 1:K \right] \right)\in {{\mathcal{R}}_{K,\alpha }}$,
\begin{align*}
 \sum\limits_{k=1}^{K}{{{w}_{k}}{{r}_{k,\alpha }}}& = \sum\limits_{k=1}^{K}{\sum\limits_{V\in {{\mathbb{V}}_{K,\alpha }}:k\in V}{{{c}_{V,\alpha }}}{{r}_{k,\alpha }}} \\
& = \sum\limits_{V\in {{\mathbb{V}}_{K,\alpha }}}{{{c}_{V,\alpha }}\sum\limits_{k\in V}{{{r}_{k,\alpha }}}} \\
& \ge \sum\limits_{V\in {{\mathbb{V}}_{K,\alpha }}}{{{c}_{V,\alpha }}{{H}_{\left| V \right|,\alpha }}}.
\end{align*}
Therefore, $\left( r_{k,\alpha }^{\left( l^{\mathbf{w}}_\alpha \right)},k\in \left[ 1:K \right] \right)$ is an optimal solution. This also shows that $\left( {{c}_{V,\alpha }},V\in {{\mathbb{V}}_{K,\alpha }} \right)$ satisfies (\ref{eq:lag_sym_f}), thus is indeed an optimal Lagrange multiplier.
\end{IEEEproof}

\subsection{Entropy Inequality}
\label{sec:ineq_sym}



Define the indicator function
\begin{align*}
\mathcal{I}\left( \text{event} \right)\triangleq\begin{cases}
   1, & \text{event is true,}  \\
   0, & \text{event is false.}
\end{cases}
\end{align*}

\begin{lemma}
\label{lem:ineq_sym_core}
For  $\mathbf{w}\in {{\mathbb{W}}_{K}}$ with ${\lambda^\mathbf{w}_{\alpha }}> 0$ and $\left( {{c}_{V,\alpha }},V\in {{\mathbb{V}}_{K,\alpha }} \right)\in {{\mathbb{C}}^\mathbf{w}_{K,\alpha }}$, define $\left( {{c}_{{V'},\alpha -1}},{V'}\in {{\mathbb{V}}_{K,\alpha -1}} \right)$ as follows:
\begin{align}
&{{c}_{\left[ 1:k \right],\alpha -1}}
\triangleq {{w}_{k}}-{{w}_{k+1}},
\quad k\in\left[1:{{l}^\mathbf{w}_{\alpha -1}}-1\right],
\nonumber \\
&{{c}_{\left[ 1:{{l}^\mathbf{w}_{\alpha -1}} \right],\alpha -1}}
\triangleq {{w}_{{{l}^\mathbf{w}_{\alpha -1}}}}-{\lambda^\mathbf{w}_{\alpha -1}},
\nonumber \\
&{{c}_{{V'},\alpha -1}} \triangleq \frac{\theta _{\alpha }^{\mathbf{w}}}{{\lambda^\mathbf{w}_{\alpha }}}\sum\limits_{V\in {{\Omega }^{\left( {{l}^\mathbf{w}_{\alpha }} \right)}_{K,\alpha }}:{V'}\subseteq V} {{c}_{V,\alpha }} + \frac{1}{{\lambda^\mathbf{w}_{\alpha }}}
\sum\limits_{V\in {{\Omega }^{\left( {{l}^\mathbf{w}_{\alpha }} \right)}_{K,\alpha }}}{{{c}_{V,\alpha }}\sum\limits_{k={{l}^\mathbf{w}_{\alpha -1}}+1}^{{{l}^\mathbf{w}_{\alpha }}}{{{c}_{\left[ 1:k \right],\alpha }}\sum\limits_{\tau =k+1}^{\alpha }{\mathcal{I}\left\{ {V'}={{\left\langle V \right\rangle }_{\left[ 1:\alpha  \right]\backslash \left\{ \tau  \right\}}} \right\}}}},\nonumber\\
&\hspace{5in} {V'}\in {{\Omega }^{\left( {{l}^\mathbf{w}_{\alpha -1}} \right)}_{K,\alpha -1}},\nonumber\\
&{{c}_{{V'},\alpha -1}}\triangleq 0, \quad \text{otherwise},
\nonumber
\end{align}
where $\theta _{\alpha }^{\mathbf{w}}$ is given by (\ref{eq:deftheta}). 
The following statements are true.

(1) $\left( {{c}_{{V'},\alpha -1}},{V'}\in {{\mathbb{V}}_{K,\alpha -1}} \right)\in {{\mathbb{C}}^\mathbf{w}_{K,\alpha -1}}$.

(2) We have
\begin{align*}
 \sum\limits_{V'\in {{\mathbb{V}}_{K,\alpha-1}}}{{{{c}_{V',\alpha-1}}H\left( {{X}_{V'}} \right)}}  \ge \sum\limits_{V\in {{\mathbb{V}}_{K,\alpha }}}{{{{c}_{V,\alpha }}H\left( {{X}_{V}} \right)}}
\end{align*}
for all $X_{\left[1:K\right]}$.
\end{lemma}
\begin{IEEEproof}
See Appendix~\ref{app:ineq_sym_core}.
\end{IEEEproof}

\begin{lemma}
\label{lem:ineq_sym}
For any $\mathbf{w}\in \mathbb{R}_{+}^{K}$, there exist $\left( {{c}_{V,\alpha }},V\in {{\mathbb{V}}_{K,\alpha }} \right)\in {{\mathbb{C}}^\mathbf{w}_{K,\alpha }}$, $\alpha \in \left[ 1:K \right]$, such that
\begin{align}
 \sum\limits_{V'\in {{\mathbb{V}}_{K,\alpha'}}}{{{{c}_{V',\alpha'}}H\left( {{X}_{V'}} \right)}}  \ge \sum\limits_{V\in {{\mathbb{V}}_{K,\alpha }}}{{{{c}_{V,\alpha }}H\left( {{X}_{V}} \right)}}\label{eq:symmetricineq}
\end{align}
for all $X_{\left[1:K\right]}$ and $\alpha\geq\alpha'$.
\end{lemma}
\begin{IEEEproof}
By symmetry, it suffices to consider $\mathbf{w}\in {{\mathbb{W}}_{K}}$. We shall first assume $w_K>0$, which implies ${\lambda^\mathbf{w}_{\alpha }}> 0$, $\alpha\in\left[1:K\right]$. Define $\left( {{c}_{V,K }},V\in {{\mathbb{V}}_{K,K }} \right)$ with
\begin{align*}
&c_{\left[1:k\right],K}\triangleq w_k-w_{k+1},\quad k\in\left[1:K\right],\\
&c_{V,K}\triangleq0,\quad\mbox{otherwise}.
\end{align*}
It is easy to verify that $\left( {{c}_{V,K }},V\in {{\mathbb{V}}_{K,K}} \right)\in {{\mathbb{C}}^\mathbf{w}_{K,K}}$. One can successively construct the desired $\left( {{c}_{V,\alpha }},V\in {{\mathbb{V}}_{K,\alpha }} \right)$ from $\alpha=K-1$ to $\alpha=1$ by invoking Lemma \ref{lem:ineq_sym_core}.

Now consider the case $w_1\geq\cdots\geq w_{K-1}>w_K=0$. The preceding argument implies the existence of $\left( {{c}'_{V,\alpha }},V\in {{\mathbb{V}}_{K,\alpha }} \right)\in {{\mathbb{C}}^\mathbf{w'}_{K-1,\alpha }}$, $\alpha \in \left[ 1:K-1 \right]$, such that
\begin{align*}
 \sum\limits_{V'\in {{\mathbb{V}}_{K-1,\alpha'}}}{{{{c}'_{V',\alpha'}}H\left( {{X}_{V'}} \right)}}  \ge \sum\limits_{V\in {{\mathbb{V}}_{K-1,\alpha }}}{{{{c}'_{V,\alpha }}H\left( {{X}_{V}} \right)}}
\end{align*}
for all $X_{\left[1:K-1\right]}$ and $\alpha\geq\alpha'$, where $\mathbf{w'}\triangleq\left(w_1,\cdots,w_{K-1}\right)$. Define $\left( {{c}_{V,\alpha }},V\in {{\mathbb{V}}_{K,\alpha }} \right)$, $\alpha\in\left[1:K-1\right]$, with
\begin{align*}
  & {{c}_{V,\alpha }}\triangleq{{c}'_{V,\alpha }},\quad K\notin V, \\
 & {{c}_{V,\alpha }}\triangleq0,\quad \mbox{otherwise},
\end{align*}
and $\left( {{c}_{V,K }},V\in {{\mathbb{V}}_{K,K}} \right)$ with
\begin{align*}
& {{c}_{V,K}}\triangleq{{c}'_{V,K-1}},\quad  K\notin V, \\
& {{c}_{V,K}}\triangleq0,\quad \mbox{otherwise}.
\end{align*}
It is easy to verify that such $\left( {{c}_{V,\alpha }},V\in {{\mathbb{V}}_{K,\alpha }} \right)$, $\alpha\in\left[1:K\right]$,
have the desired properties. The general case where $w_1\geq\cdots\geq w_{K'-1}>w_{K'}=\cdots=w_K=0$ for some $K'\le K$ can be handled via induction\footnote{If $w_1=\cdots=w_K=0$, then $c_{V,\alpha}=0$ for all $V\in {{\mathbb{V}}_{K,\alpha }}$ and $\alpha\in\left[1:K\right]$.}.
\end{IEEEproof}

\subsection{Outer Bound}
\label{sec:sym_outer}

The following result, together with (\ref{eq:innerbound}) and Lemma \ref{lem:ineq_sym}, completes the proof of Theorem \ref{th:main_sym}.
\begin{lemma}\label{lem:symmouter}
If any $\mathbf{w}\in \mathbb{R}_{+}^{K}$, there exist $\left( {{c}_{V,\alpha }},V\in {{\mathbb{V}}_{K,\alpha }} \right)\in {{\mathbb{C}}^\mathbf{w}_{K,\alpha }}$, $\alpha \in \left[ 1:K \right]$, such that (\ref{eq:symmetricineq}) holds, then 
\begin{align*}
\mathcal{R}^*_K\subseteq\mathcal{R}_K
\end{align*}
when the distribution of ${{U}_{\left[ 1:K \right],\left[ 1:K \right]}}$ is symmetrical entropy-wise.
\end{lemma}
\begin{IEEEproof}
Let $\left( {{R}_{k}}:k\in \left[ 1:K \right] \right)$ be an arbitrary admissible rate tuple. It suffices to show that
\begin{align}
\sum\limits_{k=1}^{K}{{{w}_{k}}{{R}_{k}}}\ge \sum\limits_{\alpha =1}^{K}{{\overline{{f}}^{\mathbf{w}}_{{{\alpha }}}}}.\label{eq:supportyhypersym}
\end{align}
Without loss of generality, we assume $\mathbf{w}\in {{\mathbb{W}}_{K}}$. We shall prove via induction that, when the distribution of ${{U}_{\left[ 1:K \right],\left[ 1:K \right]}}$ is symmetrical entropy-wise, for any D-MLDC system satisfying
(\ref{eq:constraint1}) and (\ref{eq:constraint2}),
\begin{align}
 \sum\limits_{k=1}^{K}{{{w}_{k}}\left( {{R}_{k}}+\epsilon  \right)}
& \ge \sum\limits_{\alpha =1}^{\beta }{{\overline{{f}}^{\mathbf{w}}_{{{\alpha }}}}}+\frac{1}{n}\sum\limits_{k=1}^{{{l}^\mathbf{w}_{\beta }}}{{{c}_{\left[ 1:k \right],\beta }}H\left( {{S}_{\left[ 1:k \right]}}|U_{\left[ 1:K \right],\left[ 1:\beta  \right]}^{n},U_{\left[ k+1:\beta  \right],K}^{n} \right)}
\nonumber \\
&\quad + \frac{1}{n}\sum\limits_{V\in \Omega _{K,\beta }^{\left( {{l}^\mathbf{w}_{\beta }} \right)}}{{{c}_{V,\beta }}H\left( {{S}_{V}}|U_{\left[ 1:K \right],\left[ 1:\beta  \right]}^{n} \right)}-\beta{{\delta }_{\epsilon }} \sum\limits_{k=1}^{K}{{{w}_{k}}} ,\quad \beta \in \left[ 1:K \right],
\label{eq:th_sym_induce}
\end{align}
where $\delta_{\epsilon}$ tends to zero as $\epsilon\rightarrow 0$. One can deduce (\ref{eq:supportyhypersym}) from (\ref{eq:th_sym_induce}) by setting $\beta=K$ and sending $\epsilon\rightarrow 0$.

The proof of (\ref{eq:th_sym_induce}) for $\beta=1$ is the same as that of (\ref{eq:induction}). Now assume that (\ref{eq:th_sym_induce}) holds for $\beta=B-1$. 
In view of (\ref{eq:symmetricineq}), we have
\begin{align}
& \sum\limits_{V'\in \Omega _{K,B-1}^{\left( {{l}^\mathbf{w}_{B-1}} \right)}}{{{c}_{V',B-1}}H\left( {{S}_{V'}}|U_{\left[ 1:K \right],\left[ 1:B-1 \right]}^{n} \right)}
\nonumber \\
& \ge \sum\limits_{k=1}^{{{l}^\mathbf{w}_{B}}}{\left( {{c}_{\left[ 1:k \right],B}}-{{c}_{\left[ 1:k \right],B-1}}\mathcal{I}\left\{ k\le {{l}^\mathbf{w}_{B-1}} \right\} \right)H\left( {{S}_{\left[ 1:k \right]}}|U_{\left[ 1:K \right],\left[ 1:B-1 \right]}^{n} \right)}
\nonumber \\
&\quad + \sum\limits_{V\in \Omega _{K,B}^{\left( {{l}^\mathbf{w}_{B}} \right)}}{{{c}_{V,B}}H\left( {{S}_{V}}|U_{\left[ 1:K \right],\left[ 1:B-1 \right]}^{n} \right)},
\label{eq:th_sym_induce1}
\end{align}
where ${{c}_{\left[ 1:k \right],B}}-{{c}_{\left[ 1:k \right],B-1}}\mathcal{I}\left\{ k\le {{l}^\mathbf{w}_{B-1}} \right\}\geq 0$, $k\in\left[1:{{l}^\mathbf{w}_{B}}\right]$, according to (\ref{eq:lag_sym_p_reprise}), (\ref{eq:moved}), and the fact that ${{c}_{\left[ 1:k \right],B-1}}={{c}_{\left[ 1:k \right],B}}$ when  $k\in\left[1: l_{B -1}^{\mathbf{w}}-1\right]$.
Moreover,
\begin{align}
& \sum\limits_{k=1}^{{{l}^\mathbf{w}_{B}}}{\left( {{c}_{\left[ 1:k \right],B}}-{{c}_{\left[ 1:k \right],B-1}}\mathcal{I}\left\{ k\le {{l}^\mathbf{w}_{B-1}} \right\} \right)H\left( {{S}_{\left[ 1:k \right]}}|U_{\left[ 1:K \right],\left[ 1:B-1 \right]}^{n} \right)}
\nonumber \\
& \ge \sum\limits_{k=1}^{{{l}^\mathbf{w}_{B}}}{\left( {{c}_{\left[ 1:k \right],B}}-{{c}_{\left[ 1:k \right],B-1}}\mathcal{I}\left\{ k\le {{l}^\mathbf{w}_{B-1}} \right\} \right)H\left( {{S}_{\left[ 1:k \right]}}|U_{\left[ 1:K \right],\left[ 1:B-1 \right]}^{n},U_{\left[ k+1:B \right],\left[ B:K \right]}^{n} \right)}
\nonumber \\
& = \sum\limits_{k=1}^{{{l}^\mathbf{w}_{B}}}{{{c}_{\left[ 1:k \right],B}}H\left( {{S}_{\left[ 1:k \right]}}|U_{\left[ 1:K \right],\left[ 1:B-1 \right]}^{n},U_{\left[ k+1:B \right],\left[ B:K \right]}^{n} \right)}
\nonumber \\
& \quad - \sum\limits_{k=1}^{{{l}^\mathbf{w}_{B-1}}}{{{c}_{\left[ 1:k \right],B-1}}H\left( {{S}_{\left[ 1:k \right]}}|U_{\left[ 1:K \right],\left[ 1:B-1 \right]}^{n},U_{\left[ k+1:B \right],\left[ B:K \right]}^{n} \right)}\nonumber \\
& \geq \sum\limits_{k=1}^{{{l}^\mathbf{w}_{B}}}{{{c}_{\left[ 1:k \right],B}}H\left( {{S}_{\left[ 1:k \right]}}|U_{\left[ 1:K \right],\left[ 1:B-1 \right]}^{n},U_{\left[ k+1:B \right],\left[ B:K \right]}^{n} \right)}
\nonumber \\
&\quad - \sum\limits_{k=1}^{{{l}^\mathbf{w}_{B-1}}}{{{c}_{\left[ 1:k \right],B-1}}H\left( {{S}_{\left[ 1:k \right]}}|U_{\left[ 1:K \right],\left[ 1:B-1 \right]}^{n},U_{\left[ k+1:B-1 \right],\left[ B-1:K \right]}^{n} \right)}.
\label{eq:th_sym_induce2}
\end{align}
Combining (\ref{eq:th_sym_induce1}) and (\ref{eq:th_sym_induce2}) gives
\begin{align}
& \sum\limits_{k=1}^{{{l}^\mathbf{w}_{B-1}}}{{{c}_{\left[ 1:k \right],B-1}}H\left( {{S}_{\left[ 1:k \right]}}|U_{\left[ 1:K \right],\left[ 1:B-1 \right]}^{n},U_{\left[ k+1:B-1 \right],\left[ B-1:K \right]}^{n} \right)}+\sum\limits_{V'\in \Omega _{K,B-1}^{\left( {{l}^\mathbf{w}_{B-1}} \right)}}{{{c}_{V',B-1}}H\left( {{S}_{V'}}|U_{\left[ 1:K \right],\left[ 1:B-1 \right]}^{n} \right)}
\nonumber \\
& \ge \sum\limits_{k=1}^{{{l}^\mathbf{w}_{B}}}{{{c}_{\left[ 1:k \right],B}}H\left( {{S}_{\left[ 1:k \right]}}|U_{\left[ 1:K \right],\left[ 1:B-1 \right]}^{n},U_{\left[ k+1:B \right],\left[ B:K \right]}^{n} \right)}+\sum\limits_{V\in \Omega _{K,B}^{\left( {{l}^\mathbf{w}_{B}} \right)}}{{{c}_{V,B}}H\left( {{S}_{V}}|U_{\left[ 1:K \right],\left[ 1:B-1 \right]}^{n} \right)}.\label{eq:tbconti}
\end{align}
Note that
\begin{align}
&H\left( {{S}_{\left[ 1:k \right]}}|U_{\left[ 1:K \right],\left[ 1:B-1 \right]}^{n},U_{\left[ k+1:B \right],\left[ B:K \right]}^{n} \right)\nonumber\\
&=H\left( {{S}_{\left[ 1:k \right]}}|U_{\left[ 1:K \right],\left[ 1:B-1 \right]}^{n},U_{\left[ k+1:B \right],\left[ B:K \right]}^{n}, {{S}_{\left[ k+1:B \right]}} \right)\nonumber\\
&=H\left( {{S}_{\left[ 1:B \right]}}|U_{\left[ 1:K \right],\left[ 1:B-1 \right]}^{n},U_{\left[ k+1:B \right],\left[ B:K \right]}^{n}, {{S}_{\left[ k+1:B \right]}} \right)\nonumber\\
&= H\left(U_{\left[ 1:B \right],\left[ 1:B \right]}^{n}, {{S}_{\left[ 1:B \right]}}|U_{\left[ 1:K \right],\left[ 1:B-1 \right]}^{n},U_{\left[ k+1:B \right],\left[ B:K \right]}^{n}, {{S}_{\left[ k+1:B \right]}} \right)\nonumber\\
&\quad-H\left(U_{\left[ 1:B \right],\left[ 1:B \right]}^{n}|U_{\left[ 1:K \right],\left[ 1:B-1 \right]}^{n},U_{\left[ k+1:B \right],\left[ B:K \right]}^{n}, {{S}_{\left[ 1:B \right]}} \right)\nonumber\\
&\ge  H\left(U_{\left[ 1:B \right],\left[ 1:B \right]}^{n}, {{S}_{\left[ 1:B \right]}}|U_{\left[ 1:K \right],\left[ 1:B-1 \right]}^{n},U_{\left[ k+1:B \right],\left[ B:K \right]}^{n}, {{S}_{\left[ k+1:B \right]}} \right)-n{{\delta }_{\epsilon }}\label{eq:invokeFanoagain}\\
&=H\left(U_{\left[ 1:B \right],\left[ 1:B \right]}^{n}|U_{\left[ 1:K \right],\left[ 1:B-1 \right]}^{n},U_{\left[ k+1:B \right],\left[ B:K \right]}^{n}, {{S}_{\left[ k+1:B \right]}} \right)\nonumber\\
&\quad+H\left({{S}_{\left[ 1:B \right]}}|U_{\left[ 1:K \right],\left[ 1:B-1 \right]}^{n},U_{\left[ 1:B \right],\left[ 1:B \right]}^{n},U_{\left[ k+1:B \right],\left[ B:K \right]}^{n}, {{S}_{\left[ k+1:B \right]}} \right)-n{{\delta }_{\epsilon }}\nonumber\\
&=H\left(U_{\left[ 1:B \right],\left[ 1:B \right]}^{n}|U_{\left[ 1:K \right],\left[ 1:B-1 \right]}^{n},U_{\left[ k+1:B \right],\left[ B:K \right]}^{n}\right)\nonumber\\
&\quad+H\left({{S}_{\left[ 1:k \right]}}|U_{\left[ 1:K \right],\left[ 1:B-1 \right]}^{n},U_{\left[ 1:B \right],\left[ 1:B \right]}^{n},U_{\left[ k+1:B \right],\left[ B:K \right]}^{n}\right)-n{{\delta }_{\epsilon }}\nonumber\\
&\geq nH_{k,B}+H\left({{S}_{\left[ 1:k \right]}}|U_{\left[ 1:K \right],\left[ 1:B \right]}^{n}, U_{\left[ k+1:B \right],\left[ B:K \right]}^{n}\right)-n{{\delta }_{\epsilon }},\quad k\in\left[1:{{l}^\mathbf{w}_{B}}\right],\label{eq:tbsub1}
\end{align}
where (\ref{eq:invokeFanoagain}) follows by (\ref{eq:constraint2}) and Fano's inequality.
Similarly, we have
\begin{align}
&H\left( {{S}_{V}}|U_{\left[ 1:K \right],\left[ 1:B-1 \right]}^{n} \right)\nonumber\\
&=H\left( U_{V,\left[ 1:B \right]}^{n},{{S}_{V}}|U_{\left[ 1:K \right],\left[ 1:B-1 \right]}^{n} \right)-H\left( U_{V,\left[ 1:\alpha  \right]}^{n}|U_{\left[ 1:K \right],\left[ 1:B-1 \right]}^{n},{{S}_{V}} \right)\nonumber\\
&\ge H\left( U_{V,\left[ 1:B  \right]}^{n},{{S}_{V}}|U_{\left[ 1:K \right],\left[ 1:B-1 \right]}^{n} \right)-n{{\delta }_{\epsilon }}\nonumber\\
&=H\left( U_{V,\left[ 1:B  \right]}^{n}|U_{\left[ 1:K \right],\left[ 1:B-1 \right]}^{n} \right)+H\left( {{S}_{V}}|U_{\left[ 1:K \right],\left[ 1:B-1 \right]}^{n},U_{V,\left[ 1:B  \right]}^{n} \right)-n{{\delta }_{\epsilon }}\nonumber\\
&\ge nH_{\left|V\right|,B}+H\left( {{S}_{V}}|U_{\left[ 1:K \right],\left[ 1:B \right]}^{n}\right)-n{{\delta }_{\epsilon }},\quad V\in \Omega _{K,B}^{\left( {{l}^\mathbf{w}_{B}} \right)}.\label{eq:tbsub2}
\end{align}
Continuing from (\ref{eq:tbconti}), 
\begin{align}
& \sum\limits_{k=1}^{{{l}^\mathbf{w}_{B-1}}}{{{c}_{\left[ 1:k \right],B-1}}H\left( {{S}_{\left[ 1:k \right]}}|U_{\left[ 1:K \right],\left[ 1:B-1 \right]}^{n},U_{\left[ k+1:B-1 \right],\left[ B-1:K \right]}^{n} \right)}+\sum\limits_{V'\in \Omega _{K,B-1}^{\left( {{l}^\mathbf{w}_{B-1}} \right)}}{{{c}_{V',B-1}}H\left( {{S}_{V'}}|U_{\left[ 1:K \right],\left[ 1:B-1 \right]}^{n} \right)}
\nonumber \\
& \ge  n\sum\limits_{V\in {{\mathbb{V}}_{K,B}}}{{{c}_{V,B}}}H_{\left|V\right|,B} + \sum\limits_{k=1}^{{{l}^\mathbf{w}_{B}}}{{{c}_{\left[ 1:k \right],B}}\cdot H\left( {{S}_{\left[ 1:k \right]}}|U_{\left[ 1:K \right],\left[ 1:B \right]}^{n},U_{\left[ k+1:B \right],\left[ B:K \right]}^{n} \right)}
\nonumber \\
&\quad+\sum\limits_{V\in \Omega _{K,B}^{\left( {{l}^\mathbf{w}_{B}} \right)}}{{{c}_{V,B}}H\left( {{S}_{V}}|U_{\left[ 1:K \right],\left[ 1:B \right]}^{n} \right)}
-n{{\delta }_{\epsilon }} \sum\limits_{V\in {{\mathbb{V}}_{K,B}}}{{{c}_{V,B}}}\label{eq:invoketwoexpan}\\
&\geq n{{\overline{{f}}^{\mathbf{w}}_{{{B}}}}}+ \sum\limits_{k=1}^{{{l}^\mathbf{w}_{B}}}{{{c}_{\left[ 1:k \right],B}}\cdot H\left( {{S}_{\left[ 1:k \right]}}|U_{\left[ 1:K \right],\left[ 1:B \right]}^{n},U_{\left[ k+1:B \right],\left[ B:K \right]}^{n} \right)}\nonumber\\
&\quad+\sum\limits_{V\in \Omega _{K,B}^{\left( {{l}^\mathbf{w}_{B}} \right)}}{{{c}_{V,B}}H\left( {{S}_{V}}|U_{\left[ 1:K \right],\left[ 1:B \right]}^{n} \right)}
-n{{\delta }_{\epsilon }} \sum\limits_{k=1}^{K}{{{w}_{k}}},
\label{eq:th_sym_induce_core2}
\end{align}
where (\ref{eq:invoketwoexpan}) is due to (\ref{eq:tbsub1}) and (\ref{eq:tbsub2}), and (\ref{eq:th_sym_induce_core2}) is due to (\ref{eq:lag_sym_f}), (\ref{eq:full_sum}) as well as Lemma \ref{lem:lp_sym}.
Combining (\ref{eq:th_sym_induce_core2}) and the induction hypothesis proves (\ref{eq:th_sym_induce}) for $\beta=B$.
\end{IEEEproof}

\section{Conclusion}
\label{sec:conclude}

We have characterized the admissible rate region of D-MLDC for the case $K\leq 3$ and the case where the source distribution is symmetrical entropy-wise. In view of the intimate connection between MLDC and its lossy counterpart known as multiple description coding \cite{tian2009}, it is expected that the results in the present work may shed new light on the robust distributed source coding problem (which is the lossy counterpart of D-MLDC) studied in \cite{chen08}.

\appendices

\section{Proof of Lemma \ref{lem:psi}}\label{app:psi}

\begin{IEEEproof}[Proof of Part (1) of Lemma \ref{lem:psi}]
 It follows by the definition of ${{\psi }_{\{i,j\}}}$ that
\begin{align*}
  {{\psi }_{\{i\}}}+{{\psi }_{\{i,j\}}}+{{\psi }_{\{j\}}}  =H\left( {{U}_{i,2}},{{U}_{j,2}} \right)
\end{align*}
when $H\left( {{U}_{i,2}},{{U}_{j,2}} \right)\ge {{\psi }_{\{i\}}}+{{\psi }_{\{j\}}}$.

When $H\left( {{U}_{i,2}},{{U}_{j,2}} \right)<{{\psi }_{\{i\}}}+{{\psi }_{\{j\}}}$, we must have
\begin{align*}
&{{\psi }_{\{i\}}}=H\left( {{U}_{i,2}}|{{U}_{k,2}} \right), \\
& {{\psi }_{\{j\}}}=H\left( {{U}_{j,2}}|{{U}_{k,2}} \right), \\
& {{\psi }_{\{i,j\}}}=0,
\end{align*}
and consequently
\begin{align*}
{{\psi }_{\{i\}}}+{{\psi }_{\{i,j\}}}+{{\psi }_{\{j\}}}=H\left( {{U}_{i,2}}|{{U}_{k,2}} \right)+H\left( {{U}_{j,2}}|{{U}_{k,2}} \right).
\end{align*}
\end{IEEEproof}

\begin{IEEEproof}[Proof of Part (2) of Lemma \ref{lem:psi}]
 Note that ${{\psi }_{\{i,j\}}}>{{\psi }_{\{i,k\}}}+{{\psi }_{\{j,k\}}}$ implies ${{\psi }_{\{i,j\}}}>0$. It then follows by the definition of ${{\psi }_{\{i,j\}}}$ that
\begin{align*}
{{\psi }_{\{i\}}}+{{\psi }_{\{i,j\}}}+{{\psi }_{\{j\}}}=H\left( {{U}_{i,2}},{{U}_{j,2}} \right).
\end{align*}
\end{IEEEproof}

\begin{IEEEproof}[Proof of Part (3) of Lemma \ref{lem:psi}]
In view of the fact that ${{\psi }_{\{i\}}}\geq H\left( {{U}_{i,2}}|{{U}_{k,2}} \right)$, ${{\psi }_{\{j\}}}\geq H\left( {{U}_{j,2}}|{{U}_{k,2}} \right)$, and ${{\psi }_{\{i,j\}}}\geq0$,
we must have
\begin{align}
&{{\psi }_{\{i\}}}=H\left( {{U}_{i,2}}|{{U}_{k,2}} \right), \label{eq:valuei}\\
&{{\psi }_{\{j\}}}=H\left( {{U}_{j,2}}|{{U}_{k,2}} \right),\nonumber\\
&{{\psi }_{\{i,j\}}}=0\nonumber
\end{align}
when ${{\psi }_{\{i\}}}+{{\psi }_{\{i,j\}}}+{{\psi }_{\{j\}}}=H\left( {{U}_{i,2}}|{{U}_{k,2}} \right)+H\left( {{U}_{j,2}}|{{U}_{k,2}} \right)$.
Note that
\begin{align}
H(U_{i,2},U_{k,2})&=H(U_{i,2}|U_{k,2})+H(U_{k,2})\nonumber\\
&={{\psi }_{\{i\}}}+H(U_{k,2})\label{eq:duei}\\
&\geq {{\psi }_{\{i\}}}+{{\psi }_{\{k\}}},\nonumber
\end{align}
where (\ref{eq:duei}) is due to (\ref{eq:valuei}). It follows by symmetry that $H\left(U_{j,2},U_{k,2}\right)\geq {{\psi }_{\{j\}}}+{{\psi }_{\{k\}}}$.
Invoking the definition of ${{\psi }_{\{i,k\}}}$ and  ${{\psi }_{\{j,k\}}}$ completes the proof of Lemma \ref{lem:psi}.
\end{IEEEproof}

\section{Proof of (\ref{eq:rightinequality}) in Lemma~\ref{lem:ineq3}}
\label{app:rightinequality}

It suffices to consider $\mathbf{w}\in{\mathbb{\mathbb{W} }_{3}}$.


Case 1A ($\mathbf{w}\in {{\mathbb{W}}^{(0)}_{3,2}}$):
\begin{align}
 & \frac{1}{2}\left( {{w}_{1}}+{{w}_{2}}-{{w}_{3}} \right)H\left( {{X}_{1}},{{X}_{2}} \right)+\frac{1}{2}\left( {{w}_{1}}-{{w}_{2}}+{{w}_{3}} \right)H\left( {{X}_{1}},{{X}_{3}} \right)  +\frac{1}{2}\left( -{{w}_{1}}+{{w}_{2}}+{{w}_{3}} \right)H\left( {{X}_{2}},{{X}_{3}} \right)\nonumber \\
  &=\left( {{w}_{1}}-{{w}_{2}} \right)\left( H\left( {{X}_{1}},{{X}_{2}} \right)+H\left( {{X}_{1}},{{X}_{3}} \right)-H\left( {{X}_{1}},{{X}_{2}},{{X}_{3}} \right) \right) +\left( {{w}_{2}}-{{w}_{3}} \right)H\left( {{X}_{1}},{{X}_{2}} \right) \nonumber\\
 &\quad +\frac{1}{2}\left( -{{w}_{1}}+{{w}_{2}}+{{w}_{3}} \right)\left( H\left( {{X}_{1}},{{X}_{2}} \right)+H\left( {{X}_{1}},{{X}_{3}} \right)+H\left( {{X}_{2}},{{X}_{3}} \right) \right)  +\left( {{w}_{1}}-{{w}_{2}} \right)H\left( {{X}_{1}},{{X}_{2}},{{X}_{3}} \right) \nonumber\\
  &\geq \left( {{w}_{1}}-{{w}_{2}} \right)H\left( {{X}_{1}} \right)+\left( {{w}_{2}}-{{w}_{3}} \right)H\left( {{X}_{1}},{{X}_{2}} \right)\nonumber\\
  &\quad +\frac{1}{2}\left( -{{w}_{1}}+{{w}_{2}}+{{w}_{3}} \right)\left( H\left( {{X}_{1}},{{X}_{2}} \right)+H\left( {{X}_{1}},{{X}_{3}} \right)+H\left( {{X}_{2}},{{X}_{3}} \right) \right)  +\left( {{w}_{1}}-{{w}_{2}} \right)H\left( {{X}_{1}},{{X}_{2}},{{X}_{3}} \right) \nonumber\\
  &\ge \left( {{w}_{1}}-{{w}_{2}} \right)H\left( {{X}_{1}} \right)+\left( {{w}_{2}}-{{w}_{3}} \right)H\left( {{X}_{1}},{{X}_{2}} \right)+{{w}_{3}}H\left( {{X}_{1}},{{X}_{2}},{{X}_{3}}\right)\label{eq:dueHan}\\
  &\geq \left( {{w}_{1}}-{{w}_{2}} \right)H\left( {{X}_{1}}|X_2,X_3 \right)+\left( {{w}_{2}}-{{w}_{3}} \right)H\left( {{X}_{1}},{{X}_{2}}|X_3 \right)+{{w}_{3}}H\left( {{X}_{1}},{{X}_{2}},{{X}_{3}}\right),\nonumber
\end{align}
where (\ref{eq:dueHan}) is due to Han's inequality~\cite{han1978}.

Case 1B ($\mathbf{w}\in {{\mathbb{W}}^{(0)}_{3,2}}$ and $\nu_2=\nu_3=1$):
\begin{align*}
 &\frac{1}{2}\left( -{{w}_{1}}+{{w}_{2}}+{{w}_{3}} \right)\left( H\left( {{X}_{2}}|{{X}_{1}}\right)+H\left( {{X}_{3}}|{{X}_{1}} \right) \right)\\
 & + \frac{1}{2}\left( {{w}_{1}}+{{w}_{2}}-{{w}_{3}} \right)H\left( {{X}_{1}},{{X}_{2}} \right)+\frac{1}{2}\left( {{w}_{1}}-{{w}_{2}}+{{w}_{3}} \right)H\left( {{X}_{1}},{{X}_{3}} \right)  \\
 &\geq \frac{1}{2}\left( -{{w}_{1}}+{{w}_{2}}+{{w}_{3}} \right)\left( H\left( {{X}_{2}}|{{X}_{1}},{{X}_{3}} \right)+H\left( {{X}_{3}}|{{X}_{1}},{{X}_{2}} \right) \right)  \\
 &\quad+ \frac{1}{2}\left( {{w}_{1}}+{{w}_{2}}-{{w}_{3}} \right)H\left( {{X}_{1}},{{X}_{2}} \right)+\frac{1}{2}\left( {{w}_{1}}-{{w}_{2}}+{{w}_{3}} \right)H\left( {{X}_{1}},{{X}_{3}} \right)\\
 &=\left( {{w}_{1}}-{{w}_{2}} \right)\left(H\left( {{X}_{1}},{{X}_{2}} \right)+H\left( {{X}_{1}},{{X}_{3}} \right)-H\left( {{X}_{1}},{{X}_{2}},{{X}_{3}} \right) \right) +\left( {{w}_{2}}-{{w}_{3}} \right)H\left( {{X}_{1}},{{X}_{2}} \right) \\
 &\quad +\frac{1}{2}\left( -{{w}_{1}}+{{w}_{2}}+{{w}_{3}} \right)\left( H\left( {{X}_{3}}|{{X}_{1}},{{X}_{2}} \right)+H\left( {{X}_{1}},{{X}_{2}} \right) \right) \\
 &\quad +\frac{1}{2}\left( -{{w}_{1}}+{{w}_{2}}+{{w}_{3}} \right)\left( H\left( {{X}_{2}}|{{X}_{1}},{{X}_{3}} \right)+H\left( {{X}_{1}},{{X}_{3}} \right)\right)  +\left( {{w}_{1}}-{{w}_{2}} \right)H\left( {{X}_{1}},{{X}_{2}},{{X}_{3}} \right) \\
 &\ge  \left( {{w}_{1}}-{{w}_{2}} \right)H\left( {{X}_{1}} \right)+\left( {{w}_{2}}-{{w}_{3}} \right)H\left( {{X}_{1}},{{X}_{2}} \right)+{{w}_{3}}H\left( {{X}_{1}},{{X}_{2}},{{X}_{3}} \right)\\
 &\geq \left( {{w}_{1}}-{{w}_{2}} \right)H\left( {{X}_{1}}|X_2,X_3 \right)+\left( {{w}_{2}}-{{w}_{3}} \right)H\left( {{X}_{1}},{{X}_{2}}|X_3 \right)+{{w}_{3}}H\left( {{X}_{1}},{{X}_{2}},{{X}_{3}} \right).
\end{align*}

Case 1C ($\nu_1=\nu_3=2$):
\begin{align*}
& \frac{1}{2}\left( {{w}_{1}}-{{w}_{2}}+{{w}_{3}} \right)\left( H\left( {{X}_{1}}|{{X}_{2}}\right)+H\left( {{X}_{3}}|{{X}_{2}} \right) \right)+\frac{1}{2}\left( {{w}_{1}}+{{w}_{2}}-{{w}_{3}} \right)H\left( {{X}_{1}},{{X}_{2}} \right)  \\
&+\frac{1}{2}\left( -{{w}_{1}}+{{w}_{2}}+{{w}_{3}} \right)H\left( {{X}_{2}},{{X}_{3}} \right) \\
 &\geq\frac{1}{2}\left( {{w}_{1}}-{{w}_{2}}+{{w}_{3}} \right)\left( H\left( {{X}_{1}}|{{X}_{2}},{{X}_{3}} \right)+H\left( {{X}_{3}}|{{X}_{1}},{{X}_{2}} \right) \right)+\frac{1}{2}\left( {{w}_{1}}+{{w}_{2}}-{{w}_{3}} \right)H\left( {{X}_{1}},{{X}_{2}} \right)   \\
 &\quad +\frac{1}{2}\left( -{{w}_{1}}+{{w}_{2}}+{{w}_{3}} \right)H\left( {{X}_{2}},{{X}_{3}} \right) \\
 &=\left( {{w}_{1}}-{{w}_{2}} \right)H\left( {{X}_{1}}|{{X}_{2}},{{X}_{3}} \right)  +\left( {{w}_{2}}-{{w}_{3}} \right)H\left( {{X}_{1}},{{X}_{2}} \right)\\
 &\quad+\frac{1}{2}\left(-w_1+w_2+w_3\right)\left(H\left(X_1|X_2,X_3\right)+H\left(X_2,X_3\right)\right)\\
 &\quad+\frac{1}{2}\left(w_1-w_2+w_3\right)\left(H\left( {{X}_{3}}|{{X}_{1}},{{X}_{2}} \right)+H\left(X_1,X_2\right)\right) \\
  &=\left( {{w}_{1}}-{{w}_{2}} \right)H\left( {{X}_{1}}|{{X}_{2}},{{X}_{3}} \right)  +\left( {{w}_{2}}-{{w}_{3}} \right)H\left( {{X}_{1}},{{X}_{2}} \right)+{{w}_{3}}H\left( {{X}_{1}},{{X}_{2}},{{X}_{3}} \right) \\
 &\ge \left( {{w}_{1}}-{{w}_{2}} \right)H\left( {{X}_{1}}|{{X}_{2}},{{X}_{3}} \right)+\left( {{w}_{2}}-{{w}_{3}} \right)H\left( {{X}_{1}},{{X}_{2}}|X_3 \right)+{{w}_{3}}H\left( {{X}_{1}},{{X}_{2}},{{X}_{3}} \right).
\end{align*}

Case 1D ($\nu_1=\nu_2=3$):
\begin{align*}
& \frac{1}{2}\left( {{w}_{1}}+{{w}_{2}}-{{w}_{3}} \right)\left( H\left( {{X}_{1}}|{{X}_{3}} \right)+H\left( {{X}_{2}}|{{X}_{3}} \right) \right)+\frac{1}{2}\left( {{w}_{1}}-{{w}_{2}}+{{w}_{3}} \right)H\left( {{X}_{1}},{{X}_{3}} \right) \\
 & +\frac{1}{2}\left( -{{w}_{1}}+{{w}_{2}}+{{w}_{3}} \right)H\left( {{X}_{2}},{{X}_{3}} \right) \\
 &\ge \frac{1}{2}\left( {{w}_{1}}+{{w}_{2}}-{{w}_{3}} \right)\left( H\left( {{X}_{1}}|{{X}_{2}},{{X}_{3}} \right)+H\left( {{X}_{2}}|{{X}_{3}} \right) \right)+\frac{1}{2}\left( {{w}_{1}}-{{w}_{2}}+{{w}_{3}} \right)H\left( {{X}_{1}},{{X}_{3}} \right) \\
 &\quad +\frac{1}{2}\left( -{{w}_{1}}+{{w}_{2}}+{{w}_{3}} \right)H\left( {{X}_{2}},{{X}_{3}} \right) \\
 & \ge\left( {{w}_{1}}-{{w}_{2}} \right)H\left( {{X}_{1}}|{{X}_{2}},{{X}_{3}} \right)+\left( {{w}_{2}}-{{w}_{3}} \right)H\left( {{X}_{1}},{{X}_{2}}|{{X}_{3}} \right) \\
 &\quad +\frac{1}{2}\left( -{{w}_{1}}+{{w}_{2}}+{{w}_{3}} \right)\left( H\left( {{X}_{1}}|{{X}_{2}},{{X}_{3}} \right)+H\left( {{X}_{2}},{{X}_{3}} \right) \right) \\
 &\quad +\frac{1}{2}\left( {{w}_{1}}-{{w}_{2}}+{{w}_{3}} \right)\left( H\left( {{X}_{2}}|{{X}_{1}},{{X}_{3}} \right)+H\left( {{X}_{1}},{{X}_{3}} \right) \right) \\
  &= \left( {{w}_{1}}-{{w}_{2}} \right)H\left( {{X}_{1}}|{{X}_{2}},{{X}_{3}} \right)+\left( {{w}_{2}}-{{w}_{3}} \right)H\left( {{X}_{1}},{{X}_{2}}|{{X}_{3}} \right)+{{w}_{3}}H\left( {{X}_{1}},{{X}_{2}},{{X}_{3}} \right).
\end{align*}

Cases 2A and 2B ($\mathbf{w}\in {{\mathbb{W}}^{(0)}_{3,2}}$):
\begin{align*}
& \left( -{{w}_{1}}+{{w}_{2}}+{{w}_{3}} \right)H\left( {{X}_{3}}|{{X}_{\nu_3}} \right)+{{w}_{2}}H\left( {{X}_{1}},{{X}_{2}} \right)+\left( {{w}_{1}}-{{w}_{2}} \right)H\left( {{X}_{1}},{{X}_{3}} \right) \\
 &\ge \left( -{{w}_{1}}+{{w}_{2}}+{{w}_{3}} \right)H\left( {{X}_{3}}|{{X}_{1}},{{X}_{2}} \right)+{{w}_{2}}H\left( {{X}_{1}},{{X}_{2}} \right)+\left( {{w}_{1}}-{{w}_{2}} \right)H\left( {{X}_{1}},{{X}_{3}} \right) \\
 & =\left( {{w}_{1}}-{{w}_{2}} \right)\left( H\left( {{X}_{1}},{{X}_{2}} \right)+H\left( {{X}_{1}},{{X}_{3}} \right)-H\left( {{X}_{1}},{{X}_{2}},{{X}_{3}} \right) \right)  +\left( {{w}_{2}}-{{w}_{3}} \right)H\left( {{X}_{1}},{{X}_{2}} \right) \\
 &\quad +\left( -{{w}_{1}}+{{w}_{2}}+{{w}_{3}} \right)\left( H\left( {{X}_{3}}|{{X}_{1}},{{X}_{2}} \right)+H\left( {{X}_{1}},{{X}_{2}} \right) \right) +\left( {{w}_{1}}-{{w}_{2}} \right)H\left( {{X}_{1}},{{X}_{2}},{{X}_{3}} \right) \\
  &\ge \left( {{w}_{1}}-{{w}_{2}} \right)H\left( {{X}_{1}} \right)+\left( {{w}_{2}}-{{w}_{3}} \right)H\left( {{X}_{1}},{{X}_{2}} \right)+{{w}_{3}}H\left( {{X}_{1}},{{X}_{2}},{{X}_{3}} \right)\\
  &\ge \left( {{w}_{1}}-{{w}_{2}} \right)H\left( {{X}_{1}}|X_2,X_3 \right)+\left( {{w}_{2}}-{{w}_{3}} \right)H\left( {{X}_{1}},{{X}_{2}} |X_3\right)+{{w}_{3}}H\left( {{X}_{1}},{{X}_{2}},{{X}_{3}} \right).
\end{align*}

Cases 2C and 5B:
\begin{align*}
&\left(w_1-w_2\right)H\left(X_1|X_{\nu_1}\right)+w_3H\left( {{X}_{3}}|{{X}_{\nu_3}} \right)+{{w}_{2}}H\left( {{X}_{1}},{{X}_{2}} \right)\\
&\geq\left(w_1-w_2\right)H\left(X_1|X_2,X_3\right)+w_3H\left( {{X}_{3}}|X_1,X_2\right)+{{w}_{2}}H\left( {{X}_{1}},{{X}_{2}} \right)\\
  &=\left( {{w}_{1}}-{{w}_{2}} \right)H\left( {{X}_{1}}|{{X}_{2}},{{X}_{3}} \right)  +\left( {{w}_{2}}-{{w}_{3}} \right)H\left( {{X}_{1}},{{X}_{2}} \right)+{{w}_{3}}H\left( {{X}_{1}},{{X}_{2}},{{X}_{3}} \right)\\
  &\ge\left( {{w}_{1}}-{{w}_{2}} \right)H\left( {{X}_{1}}|{{X}_{2}},{{X}_{3}} \right)+\left( {{w}_{2}}-{{w}_{3}} \right)H\left( {{X}_{1}},{{X}_{2}}|X_3 \right)+{{w}_{3}}H\left( {{X}_{1}},{{X}_{2}},{{X}_{3}} \right).
\end{align*}

Cases 3A and 3B ($\mathbf{w}\in {{\mathbb{W}}^{(0)}_{3,2}}$):
\begin{align*}
& \left( -{{w}_{1}}+{{w}_{2}}+{{w}_{3}} \right)H\left( {{X}_{2}}|{{X}_{\nu_2}} \right)+\left( {{w}_{1}}-{{w}_{3}} \right)H\left( {{X}_{1}},{{X}_{2}} \right)+{{w}_{3}}H\left( {{X}_{1}},{{X}_{3}} \right) \\
  &\geq \left( -{{w}_{1}}+{{w}_{2}}+{{w}_{3}} \right)H\left( {{X}_{2}}|{{X}_{1}},{{X}_{3}} \right)+\left( {{w}_{1}}-{{w}_{3}} \right)H\left( {{X}_{1}},{{X}_{2}} \right)+{{w}_{3}}H\left( {{X}_{1}},{{X}_{3}} \right) \\
  &=\left( {{w}_{1}}-{{w}_{2}} \right)\left( H\left( {{X}_{1}},{{X}_{2}} \right)+H\left( {{X}_{1}},{{X}_{3}} \right)-H\left( {{X}_{1}},{{X}_{2}},{{X}_{3}} \right) \right) +\left( {{w}_{2}}-{{w}_{3}} \right)H\left( {{X}_{1}},{{X}_{2}} \right) \\
 &\quad +\left( {{w}_{1}}-{{w}_{2}} \right)H\left( {{X}_{1}},{{X}_{2}},{{X}_{3}} \right)+\left( -{{w}_{1}}+{{w}_{2}}+{{w}_{3}} \right)\left( H\left( {{X}_{2}}|{{X}_{1}},{{X}_{3}} \right)+H\left( {{X}_{1}},{{X}_{3}} \right) \right) \\
  &\ge\left( {{w}_{1}}-{{w}_{2}} \right)H\left( {{X}_{1}} \right) +\left( {{w}_{2}}-{{w}_{3}} \right)H\left( {{X}_{1}},{{X}_{2}} \right) +{{w}_{3}}H\left( {{X}_{1}},{{X}_{2}},{{X}_{3}} \right) \\
  &\ge\left( {{w}_{1}}-{{w}_{2}} \right)H\left( {{X}_{1}}|X_2,X_3 \right) +\left( {{w}_{2}}-{{w}_{3}} \right)H\left( {{X}_{1}},{{X}_{2}} |X_3\right) +{{w}_{3}}H\left( {{X}_{1}},{{X}_{2}},{{X}_{3}} \right).
\end{align*}

Cases 3D and 5D ($\nu_1=\nu_2=3$):
\begin{align*}
&\left(w_1-w_3\right)H\left(X_1|X_3\right)+w_2H\left(X_2|X_3\right)+w_3H\left(X_1,X_3\right)\\
&\ge\left(w_1-w_3\right)H\left(X_1|X_3\right)+w_2H\left(X_2|X_1,X_3\right)+w_3H\left(X_1,X_3\right)\\
 &=\left( {{w}_{1}}-{{w}_{2}} \right)H\left( {{X}_{1}}|{{X}_{3}} \right) +\left( {{w}_{2}}-{{w}_{3}} \right)H\left( {{X}_{1}},{{X}_{2}}|{{X}_{3}} \right) +{{w}_{3}}H\left( {{X}_{1}},{{X}_{2}},{{X}_{3}} \right)\\
  &\ge\left( {{w}_{1}}-{{w}_{2}} \right)H\left( {{X}_{1}}|X_2,{{X}_{3}} \right) +\left( {{w}_{2}}-{{w}_{3}} \right)H\left( {{X}_{1}},{{X}_{2}}|{{X}_{3}} \right) +{{w}_{3}}H\left( {{X}_{1}},{{X}_{2}},{{X}_{3}} \right).
\end{align*}


Cases 4A and 4C:
\begin{align*}
& \left( {{w}_{1}}-{{w}_{2}}+{{w}_{3}} \right)H\left( {{X}_{1}}|{{X}_{\nu_1}} \right)+\left( {{w}_{2}}-{{w}_{3}} \right)H\left( {{X}_{1}},{{X}_{2}} \right)+{{w}_{3}}H\left( {{X}_{2}},{{X}_{3}} \right) \\
&\ge \left( {{w}_{1}}-{{w}_{2}}+{{w}_{3}} \right)H\left( {{X}_{1}}|{{X}_{2}},{{X}_{3}} \right)+\left( {{w}_{2}}-{{w}_{3}} \right)H\left( {{X}_{1}},{{X}_{2}} \right)+{{w}_{3}}H\left( {{X}_{2}},{{X}_{3}} \right) \\
& = \left( {{w}_{1}}-{{w}_{2}} \right)H\left( {{X}_{1}}|{{X}_{2}},{{X}_{3}} \right)+\left( {{w}_{2}}-{{w}_{3}} \right)H\left( {{X}_{1}},{{X}_{2}} \right) +{{w}_{3}}H\left( {{X}_{1}},{{X}_{2}},{{X}_{3}} \right) \\
& \ge \left( {{w}_{1}}-{{w}_{2}} \right)H\left( {{X}_{1}}|{{X}_{2}},{{X}_{3}} \right)+\left( {{w}_{2}}-{{w}_{3}} \right)H\left( {{X}_{1}},{{X}_{2}} |X_3\right) +{{w}_{3}}H\left( {{X}_{1}},{{X}_{2}},{{X}_{3}} \right).
\end{align*}

Case 4D ($\nu_1=\nu_2=3$):
\begin{align*}
&w_1H\left(X_1|X_3\right)+\left(w_2-w_3\right)H\left(X_2|X_3\right)+w_3H\left(X_2,X_3\right)\\
&\ge \left( {{w}_{1}}-{{w}_{2}}+{{w}_{3}} \right)H\left( {{X}_{1}}|{{X}_{2}},{{X}_{3}} \right)  +\left( {{w}_{2}}-{{w}_{3}} \right)\left( H\left( {{X}_{1}}|{{X}_{3}} \right)+H\left( {{X}_{2}}|{{X}_{3}} \right) \right)+{{w}_{3}}H\left( {{X}_{2}},{{X}_{3}} \right) \\
& = \left( {{w}_{1}}-{{w}_{2}} \right)H\left( {{X}_{1}}|{{X}_{2}},{{X}_{3}} \right) +\left( {{w}_{2}}-{{w}_{3}} \right)\left( H\left( {{X}_{1}}|{{X}_{3}} \right)+H\left( {{X}_{2}}|{{X}_{3}} \right) \right)  +{{w}_{3}}H\left( {{X}_{1}},{{X}_{2}},{{X}_{3}} \right)\\
& \ge \left( {{w}_{1}}-{{w}_{2}} \right)H\left( {{X}_{1}}|{{X}_{2}},{{X}_{3}} \right)+\left( {{w}_{2}}-{{w}_{3}} \right)H\left( {{X}_{1}},{{X}_{2}}|{{X}_{3}} \right)+{{w}_{3}}H\left( {{X}_{1}},{{X}_{2}},{{X}_{3}} \right).
\end{align*}


Cases 5A and 5C:
\begin{align*}
& \left( {{w}_{1}}-{{w}_{2}}-{{w}_{3}} \right)H\left( {{X}_{1}}|{{X}_{\nu_1}} \right)+{{w}_{2}}H\left( {{X}_{1}},{{X}_{2}} \right)+{{w}_{3}}H\left( {{X}_{1}},{{X}_{3}} \right) \\
  &\ge \left( {{w}_{1}}-{{w}_{2}} \right)H\left( {{X}_{1}}|{{X}_{2}},{{X}_{3}} \right)  +\left( {{w}_{2}}-{{w}_{3}} \right)H\left( {{X}_{1}},{{X}_{2}} \right) +{{w}_{3}}\left( H\left( {{X}_{1}},{{X}_{2}} \right)+H\left( {{X}_{1}},{{X}_{3}} \right)-H\left( {{X}_{1}} \right) \right) \\
 &\ge  \left( {{w}_{1}}-{{w}_{2}} \right)H\left( {{X}_{1}}|{{X}_{2}},{{X}_{3}} \right)+\left( {{w}_{2}}-{{w}_{3}} \right)H\left( {{X}_{1}},{{X}_{2}} \right)+{{w}_{3}}H\left( {{X}_{1}},{{X}_{2}},{{X}_{3}} \right)\\
 &\ge  \left( {{w}_{1}}-{{w}_{2}} \right)H\left( {{X}_{1}}|{{X}_{2}},{{X}_{3}} \right)+\left( {{w}_{2}}-{{w}_{3}} \right)H\left( {{X}_{1}},{{X}_{2}}|X_3 \right)+{{w}_{3}}H\left( {{X}_{1}},{{X}_{2}},{{X}_{3}} \right).
\end{align*}



\section{Proof of Lemma~\ref{lem:opt_in_region}}
\label{app:opt_in_region}

The following result is needed for the proof of Lemma~\ref{lem:opt_in_region}.

\begin{lemma}
\label{lem:sym_div_ineq}
Assume that $H\left( {{X}_{V}} \right)=H\left( {{X}_{{{V}'}}} \right)$ for all $V,{V}'\subseteq \left[ 1:K \right]$ with $|V|=|V'|$. We have
\begin{align*}
\frac{H\left( {{X}_{\left[ 1:{{i}_{1}} \right] }}|{{X}_{\left[ {{i}_{1}}+1:j \right] }} \right)}{{{i}_{1}}}\le \frac{H\left( {{X}_{\left[ 1:{{i}_{2}} \right] }}|{{X}_{\left[ {{i}_{2}}+1:j \right] }} \right)}{{{i}_{2}}}
\end{align*}
for any ${{i}_{1}},{{i}_{2}},j\in \left[ 1:K \right]$ such that ${{i}_{1}}\le {{i}_{2}}\le j$.
\end{lemma}
\begin{IEEEproof}
It suffices to consider the case $i_1<i_2$. Note that
\begin{align}
 {{i}_{2}}H\left( {{X}_{\left[ 1:{{i}_{2}} \right]}}|{{X}_{\left[ {{i}_{2}}+1:j \right]}} \right)& = \sum\limits_{k=1}^{{{i}_{2}}}{H\left( {{X}_{\left[ 1:{{i}_{2}} \right]}}|{{X}_{\left[ {{i}_{2}}+1:j \right]}} \right)} \nonumber \\
& = \sum\limits_{k=1}^{{{i}_{2}}}{H\left( {{X}_{k}}|{{X}_{\left[ {{i}_{2}}+1:j \right]}} \right)}+\sum\limits_{k=1}^{{{i}_{2}}}{H\left( {{X}_{\left[ 1:{{i}_{2}} \right]\backslash \left\{ k \right\}}}|{{X}_{\left\{ k \right\}\bigcup \left[ {{i}_{2}}+1:j \right]}} \right)} \nonumber \\
& \ge H\left( {{X}_{\left[ 1:{{i}_{2}} \right]}}|{{X}_{\left[ {{i}_{2}}+1:j \right]}} \right)+{{i}_{2}}H\left( {{X}_{\left[ 1:{{i}_{2}}-1 \right]}}|{{X}_{\left[ {{i}_{2}}:j \right]}} \right).\nonumber
\end{align}
Therefore,
\begin{align*}
\frac{H\left( {{X}_{\left[ 1:{{i}_{2}}-1 \right] }}|{{X}_{\left[ {{i}_{2}}:j \right] }} \right)}{{{i}_{2}}-1}\le \frac{H\left( {{X}_{\left[ 1:{{i}_{2}} \right] }}|{{X}_{\left[ {{i}_{2}}+1:j \right] }} \right)}{{{i}_{2}}}.
\end{align*}
One can readily complete the proof via induction.
\end{IEEEproof}

Now we are ready to prove Lemma~\ref{lem:opt_in_region}.
\begin{IEEEproof}[Proof of Lemma~\ref{lem:opt_in_region}]
Consider an arbitrary $V\in {{\mathbb{V}}_{K,\alpha }}$. Let ${{V}_{1}}\triangleq V\cap \left[ 1:l \right]$ and ${{V}_{2}}\triangleq V\backslash {{V}_{1}}$. It suffices to show that
\begin{align}
 \sum\limits_{k\in {{V}_{1}}}{H\left( {{U}_{k,\alpha }}|{{U}_{\left[ k+1:\alpha  \right],\alpha }} \right)}+\left| {{V}_{2}} \right|\frac{H\left( {{U}_{\left[ l+1:\alpha  \right],\alpha}} \right)}{\alpha -l}  \ge H\left( {{U}_{\left[ 1:\left| V \right| \right],\alpha }}|{{U}_{\left[ \left| V \right|+1:\alpha  \right],\alpha }} \right).\label{eq:sufficeregion}
\end{align}

First consider the case $V_i\neq\emptyset$, $i=1,2$. Note that
\begin{align}
\sum\limits_{k \in {{V}_{1}}}{H\left( {{U}_{k,\alpha }}|{{U}_{\left[ k+1:\alpha  \right],\alpha }} \right)}& = \sum\limits_{\tau =1}^{\left| {{V}_{1}} \right|}{H\left( {{U}_{{{\left\langle {{V}_{1}} \right\rangle }_{\tau }},\alpha }}|{{U}_{\left[ {{\left\langle {{V}_{1}} \right\rangle }_{\tau }}+1:\alpha  \right],\alpha }} \right)} \nonumber \\
& \ge \sum\limits_{\tau =1}^{\left| {{V}_{1}} \right|}{H\left( {{U}_{\tau ,\alpha }}|{{U}_{\left[ \tau +1:\alpha  \right],\alpha }} \right)} \label{eq:taubig}\\
& = H\left( {{U}_{\left[ 1:\left| {{V}_{1}} \right| \right],\alpha }}|{{U}_{\left[ \left| {{V}_{1}} \right|+1:\alpha  \right],\alpha }} \right),\label{eqn:feasible_v1}
\end{align}
where (\ref{eq:taubig}) is due to the fact that ${{\left\langle {{V}_{1}} \right\rangle }_{\tau }}\ge \tau $ for $\tau \in \left[ 1:\left| {{V}_{1}} \right| \right]$ and that the source distribution is symmetrical entropy-wise. Moreover, we have
\begin{align}
\left| {{V}_{2}} \right|\frac{H\left( {{U}_{\left[ l+1:\alpha  \right],\alpha}} \right)}{\alpha -l}&\ge \left| {{V}_{2}} \right|\frac{H\left( {{U}_{\left[ 1:\alpha -l \right],\alpha}}|{{U}_{\left[ \alpha -l+1:\alpha -\left| {{V}_{1}} \right| \right],\alpha}} \right)}{\alpha -l} \nonumber \\
& \ge H\left( {{U}_{\left[ 1:\left| {{V}_{2}} \right| \right],\alpha}}|{{U}_{\left[ \left| {{V}_{2}} \right|+1:\alpha -\left| {{V}_{1}} \right| \right],\alpha}} \right) \label{eq:invokelemma}\\
 & = H\left( {{U}_{\left[ \left| {{V}_{1}} \right|+1:\left| V \right| \right],\alpha}}|{{U}_{\left[ \left| V \right|+1:\alpha  \right],\alpha}} \right),\label{eqn:feasible_v2}
\end{align}
where (\ref{eq:invokelemma}) is due to Lemma~\ref{lem:sym_div_ineq} and the fact that $\alpha -l \ge \left| {{V}_{2}} \right|$.
Combining (\ref{eqn:feasible_v1}) and (\ref{eqn:feasible_v2}) proves (\ref{eq:sufficeregion}) for the case $V_i\neq\emptyset$, $i=1,2$.



Note that (\ref{eq:sufficeregion}) degenerates to (\ref{eqn:feasible_v1}) when $V_2=\emptyset$ and degenerates to (\ref{eqn:feasible_v2}) when $V_1=\emptyset$. This completes the proof of Lemma \ref{lem:opt_in_region}.
\end{IEEEproof}




\section{Proof of Lemma~\ref{lem:prepare_lag_sym}}
\label{app:prepare_lag_sym}

\begin{IEEEproof}[Proof of (\ref{eq:partial_sum})]
Note that
\begin{align*}
\sum\limits_{k={{l}^\mathbf{w}_{\alpha }}+1}^{K}{{{w}_{k}}}&=\sum\limits_{k={{l}^\mathbf{w}_{\alpha }}+1}^{K}\sum_{V\in {{\Omega }^{\left( {{l}^\mathbf{w}_{\alpha }} \right)}_{K,\alpha }}:k\in V}{{{c}_{V,\alpha }}}\\
& = \sum\limits_{V\in {{\Omega }^{\left( {{l}^\mathbf{w}_{\alpha }} \right)}_{K,\alpha }}}{{{c}_{V,\alpha }}\sum\limits_{k\in V\backslash \left[ 1:{{l}^\mathbf{w}_{\alpha }} \right]}{1}}
\\
& = \left( \alpha -{{l}^\mathbf{w}_{\alpha }} \right)\sum\limits_{V\in {{\Omega }^{\left( {{l}^\mathbf{w}_{\alpha }} \right)}_{K,\alpha }}}{{{c}_{V,\alpha }}},
\end{align*}
from which the desired result follows immediately.
\end{IEEEproof}

\begin{IEEEproof}[Proof of (\ref{eq:lag_sym_w_reprise})] 
 Consider the following two cases.

\emph{(Case 1)}
$k\in \left[ 1:{{l}^\mathbf{w}_{\alpha }} \right]$:
\begin{align}
\sum\limits_{V\in {{\mathbb{V}}_{K,\alpha }}:k\in V}{{{c}_{V,\alpha }}}
& =\sum\limits_{i=k}^{{{l}^\mathbf{w}_{\alpha }}}{{{c}_{\left[ 1:i \right],\alpha }}}+\sum\limits_{V\in \Omega _{K,\alpha }^{\left( {{l}^\mathbf{w}_{\alpha }} \right)}}{{{c}_{V,\alpha }}} \nonumber\\
& ={{w}_{k}}-\lambda^\mathbf{w}_{\alpha }+\sum\limits_{V\in \Omega _{K,\alpha }^{\left( {{l}^\mathbf{w}_{\alpha }} \right)}}{{{c}_{V,\alpha }}} \nonumber\\
& ={{w}_{k}},\label{eq:duetosum}
\end{align}
where (\ref{eq:duetosum}) is due to (\ref{eq:partial_sum}).

\emph{(Case 2)}
$k\in \left[ {{l}^\mathbf{w}_{\alpha }}+1:K \right]$:
\begin{align*}
\sum\limits_{V\in {{\mathbb{V}}_{K,\alpha }}:k\in V}{{{c}_{V,\alpha }}}=\sum\limits_{V\in \Omega _{K,\alpha }^{\left( {{l}^\mathbf{w}_{\alpha }} \right)}:k\in V}{{{c}_{V,\alpha }}}={{w}_{k}}.
\end{align*}
\end{IEEEproof}

\begin{IEEEproof}[Proof of (\ref{eq:lag_sym_p_reprise})]
 It suffices to verify that ${{w}_{{{l}^\mathbf{w}_{\alpha }}}}-\lambda^\mathbf{w}_{\alpha }\geq 0$ when ${l}^\mathbf{w}_{\alpha }\geq 1$. Indeed, this is a simple consequence of the fact that  $\mathbf{w}\in{{\mathbb{W}}^{\left( {{l}^\mathbf{w}_{\alpha }} \right)}_{K,\alpha }}$.
\end{IEEEproof}

\begin{IEEEproof}[Proof of (\ref{eq:moved})] 
 Consider the following three cases.

\emph{(Case 1)}
$l_{\alpha -1}^{\mathbf{w}}\le l_{\alpha }^{\mathbf{w}}-2$: 
Note that
\begin{align}
 \sum\limits_{k=l_{\alpha -1}^{\mathbf{w}}+1}^{l_{\alpha }^{\mathbf{w}}}{\left( \alpha-1 -k \right){{c}_{\left[ 1:k \right],\alpha }}}  = \left( \alpha-1 -l_{\alpha -1}^{\mathbf{w}} \right){{w}_{l_{\alpha -1}^{\mathbf{w}}+1}}-\sum\limits_{k=l_{\alpha -1}^{\mathbf{w}}+1}^{l_{\alpha }^{\mathbf{w}}}{{{w}_{k}}}-\left( \alpha-1 -l_{\alpha }^{\mathbf{w}} \right)\lambda _{\alpha }^{\mathbf{w}}\label{eq:tbslater1}
\end{align}
and
\begin{align}
\left( \alpha -l_{\alpha }^{\mathbf{w}} \right)\lambda _{\alpha }^{\mathbf{w}}+\sum\limits_{k=l_{\alpha -1}^{\mathbf{w}}+1}^{l_{\alpha }^{\mathbf{w}}}{{{w}_{k}}} 
& =\sum\limits_{k=l_{\alpha }^{\mathbf{w}}+1}^{K}{{{w}_{k}}}+\sum\limits_{k=l_{\alpha -1}^{\mathbf{w}}+1}^{l_{\alpha }^{\mathbf{w}}}{{{w}_{k}}} \nonumber\\
& =\sum\limits_{k=l_{\alpha-1 }^{\mathbf{w}}+1}^{K}{{{w}_{k}}} \nonumber\\
& =\left( \alpha -1-l_{\alpha -1}^{\mathbf{w}} \right)\lambda _{\alpha -1}^{\mathbf{w}}.\label{eq:tbslater2}
\end{align}
We have
\begin{align}
\theta _{\alpha }^{\mathbf{w}}
& = \left( \lambda _{\alpha }^{\mathbf{w}}-\sum\limits_{k=l_{\alpha -1}^{\mathbf{w}}+1}^{l_{\alpha }^{\mathbf{w}}}{\left( \alpha-1 -k \right){{c}_{\left[ 1:k \right],\alpha }}} \right)\frac{1}{\alpha-1 -l_{\alpha -1}^{\mathbf{w}}}\nonumber \\
& = \left( \lambda _{\alpha }^{\mathbf{w}}- \left( \alpha-1 -l_{\alpha -1}^{\mathbf{w}} \right){{w}_{l_{\alpha -1}^{\mathbf{w}}+1}}+\sum\limits_{k=l_{\alpha -1}^{\mathbf{w}}+1}^{l_{\alpha }^{\mathbf{w}}}{{{w}_{k}}}+\left( \alpha-1 -l_{\alpha }^{\mathbf{w}} \right)\lambda _{\alpha }^{\mathbf{w}}  \right)\frac{1}{\alpha-1 -l_{\alpha -1}^{\mathbf{w}}}\label{eq:invoketb1} \\
& = \left( \left( \alpha -l_{\alpha }^{\mathbf{w}} \right)\lambda _{\alpha }^{\mathbf{w}}+\sum\limits_{k=l_{\alpha -1}^{\mathbf{w}}+1}^{l_{\alpha }^{\mathbf{w}}}{{{w}_{k}}}-\left( \alpha-1 -l_{\alpha -1}^{\mathbf{w}} \right){{w}_{l_{\alpha -1}^{\mathbf{w}}+1}} \right)\frac{1}{\alpha-1 -l_{\alpha -1}^{\mathbf{w}}} \nonumber\\
& = \left( \left( \alpha-1 -l_{\alpha -1}^{\mathbf{w}} \right)\lambda _{\alpha -1}^{\mathbf{w}}-\left( \alpha-1 -l_{\alpha -1}^{\mathbf{w}} \right){{w}_{l_{\alpha -1}^{\mathbf{w}}+1}} \right)\frac{1}{\alpha-1 -l_{\alpha -1}^{\mathbf{w}}}\label{eq:invoketb2} \\
& = \lambda _{\alpha -1}^{\mathbf{w}}-{{w}_{l_{\alpha -1}^{\mathbf{w}}+1}}\label{eq:almost1} \\
& = {{c}_{\left[ 1:l_{\alpha -1}^{\mathbf{w}} \right],\alpha }}-{{c}_{\left[ 1:l_{\alpha -1}^{\mathbf{w}} \right],\alpha -1}},\nonumber
\end{align}
where (\ref{eq:invoketb1}) and (\ref{eq:invoketb2}) are due to (\ref{eq:tbslater1}) and (\ref{eq:tbslater2}), respectively.  It can be verified that
\begin{align}
\left(\alpha -1-l_{\alpha -1}^{\mathbf{w}}\right)\lambda _{\alpha -1}^{\mathbf{w}}&=\sum\limits_{k=l_{\alpha -1}^{\mathbf{w}}+1}^{K}{{{w}_{k}}}\nonumber\\
&=\sum\limits_{k=l_{\alpha -1}^{\mathbf{w}}+2}^{K}{{{w}_{k}}}+{{w}_{l_{\alpha -1}^{\mathbf{w}}+1}} \nonumber\\
& \ge \left( \alpha -2-l_{\alpha -1}^{\mathbf{w}} \right){{w}_{{{l}^{\mathbf{w}}_{\alpha -1}}+1}}+{{w}_{l_{\alpha -1}^{\mathbf{w}}+1}}. \label{eq:use_ww_def}\\
& =\left( \alpha -1-l_{\alpha -1}^{\mathbf{w}} \right){{w}_{{{l}^{\mathbf{w}}_{\alpha -1}}+1}},\label{eq:neverending}
\end{align}
where (\ref{eq:use_ww_def}) follows from the fact that $\mathbf{w}\in {{\mathbb{W} }^{\left( {{l}^\mathbf{w}_{\alpha -1}} \right)}_{K,\alpha -1}}$. Combining (\ref{eq:almost1}) and (\ref{eq:neverending}) proves $\theta _{\alpha }^{\mathbf{w}}\ge 0$.

\emph{(Case 2)}
$l_{\alpha -1}^{\mathbf{w}}=l_{\alpha }^{\mathbf{w}}-1$: Note that
\begin{align}
\left(\alpha -l_{\alpha }^{\mathbf{w}}\right)\lambda _{\alpha }^{\mathbf{w}}&=\sum\limits_{k=l_{\alpha }^{\mathbf{w}}+1}^{K}{{{w}_{k}}}\nonumber\\
&=\left(\alpha-1 -l_{\alpha-1 }^{\mathbf{w}}\right)\lambda _{\alpha-1 }^{\mathbf{w}}-w_{l_{\alpha }^{\mathbf{w}}}\nonumber\\
&=\left(\alpha -l_{\alpha }^{\mathbf{w}}\right)\lambda _{\alpha-1 }^{\mathbf{w}}-w_{l_{\alpha }^{\mathbf{w}}}.\label{eq:onemoretb}
\end{align}
We have
\begin{align}
\theta _{\alpha }^{\mathbf{w}}
& = \left(\lambda _{\alpha }^{\mathbf{w}}-{\left( \alpha-1 -l_{\alpha}^{\mathbf{w}} \right){{c}_{\left[ 1:l_{\alpha}^{\mathbf{w}} \right],\alpha }}} \right)\frac{1}{\alpha-1 -l_{\alpha -1}^{\mathbf{w}}}\nonumber \\
&= \left(\lambda _{\alpha }^{\mathbf{w}}-{\left( \alpha-1 -l_{\alpha}^{\mathbf{w}} \right)\left({{w}_{l_{\alpha}^{\mathbf{w}}}}-\lambda _{\alpha}^{\mathbf{w}}\right)} \right)\frac{1}{\alpha -l_{\alpha }^{\mathbf{w}}}\nonumber \\
& = \lambda _{\alpha -1}^{\mathbf{w}}-{{w}_{l_{\alpha}^{\mathbf{w}}}}\label{eq:duetoonemoretb}\\
& = {{c}_{\left[ 1:l_{\alpha -1}^{\mathbf{w}} \right],\alpha }}-{{c}_{\left[ 1:l_{\alpha -1}^{\mathbf{w}} \right],\alpha -1}},\nonumber
\end{align}
where (\ref{eq:duetoonemoretb}) is due to (\ref{eq:onemoretb}). The fact that $\theta _{\alpha }^{\mathbf{w}}\ge 0$ follows by (\ref{eq:neverending}) and (\ref{eq:duetoonemoretb}).

\emph{(Case 3)}
$
l_{\alpha -1}^{\mathbf{w}}=l_{\alpha }^{\mathbf{w}}
$: Note that
\begin{align*}
\theta _{\alpha }^{\mathbf{w}}
& =\frac{1}{\alpha-1 -l_{\alpha -1}^{\mathbf{w}}}\lambda _{\alpha }^{\mathbf{w}}\ge 0.
\end{align*}
Moreover, we have
\begin{align*}
\frac{1}{\alpha-1 -l_{\alpha -1}^{\mathbf{w}}}\lambda _{\alpha }^{\mathbf{w}}& = \frac{1}{\left(\alpha -1-l_{\alpha }^{\mathbf{w}}\right)\left(\alpha -l_{\alpha }^{\mathbf{w}}\right)}\sum\limits_{k=l_{\alpha }^{\mathbf{w}}+1}^{K}{{{w}_{k}}} \\
& = \left( \frac{1}{\alpha -1-l_{\alpha }^{\mathbf{w}}}-\frac{1}{\alpha -l_{\alpha }^{\mathbf{w}}} \right)\sum\limits_{k=l_{\alpha }^{\mathbf{w}}+1}^{K}{{{w}_{k}}} \\
& = \lambda _{\alpha -1}^{\mathbf{w}}-\lambda _{\alpha }^{\mathbf{w}} \\
& = {{c}_{\left[ 1:l_{\alpha -1}^{\mathbf{w}} \right],\alpha }}-{{c}_{\left[ 1:l_{\alpha -1}^{\mathbf{w}} \right],\alpha -1}}.
\end{align*}

\end{IEEEproof}

\begin{IEEEproof}[Proof of (\ref{eq:full_sum})]
Consider the following two cases.

\emph{(Case 1)}
$l^\mathbf{w}_{\alpha}=0$:
\begin{align}
\sum\limits_{V\in {{\mathbb{V}}_{K,\alpha }}}{{{c}_{V,\alpha }}}&=\sum\limits_{V\in \Omega _{K,\alpha }^{\left( 0 \right)}}{{{c}_{V,\alpha }}}\nonumber\\
&=\frac{1}{\alpha }\sum\limits_{k=1}^{K}{{{w}_{k}}},\label{eq:invokepart1}
\end{align}
where (\ref{eq:invokepart1}) is due to (\ref{eq:partial_sum}).

\emph{(Case 2)}
$l^\mathbf{w}_{\alpha}>0$:
\begin{align}
\sum\limits_{V\in {{\mathbb{V}}_{K,\alpha }}}{{{c}_{V,\alpha }}} &= \sum\limits_{k=1}^{l^\mathbf{w}_{\alpha}-1}{{{c}_{\left[ 1:k \right],\alpha }}}+{{c}_{\left[ 1:{l^\mathbf{w}_{\alpha}} \right],\alpha }}+\sum\limits_{V\in \Omega _{K,\alpha }^{\left( l^\mathbf{w}_{\alpha} \right)}}{{{c}_{V,\alpha }}} \nonumber\\
& = \sum\limits_{k=1}^{l^\mathbf{w}_{\alpha}-1}{\left( {{w}_{k}}-{{w}_{k+1}} \right)}+\left( {{w}_{l^\mathbf{w}_{\alpha}}}-\lambda^\mathbf{w}_{\alpha} \right)+\sum\limits_{V\in \Omega _{K,\alpha }^{\left( l^\mathbf{w}_{\alpha} \right)}}{{{c}_{V,\alpha }}} \nonumber\\
&=\sum\limits_{k=1}^{l^\mathbf{w}_{\alpha}-1}{\left( {{w}_{k}}-{{w}_{k+1}} \right)}+\left( {{w}_{l^\mathbf{w}_{\alpha}}}-\lambda^\mathbf{w}_{\alpha} \right)+\lambda^\mathbf{w}_{\alpha}\label{eq:invokepart2}\\
& = {{w}_{1}},\nonumber
\end{align}
where (\ref{eq:invokepart2}) is due to (\ref{eq:partial_sum}).
\end{IEEEproof}

%

\section{Proof of Lemma~\ref{lem:ineq_sym_core}}
\label{app:ineq_sym_core}

\begin{IEEEproof}[Proof of Part (1) of Lemma~\ref{lem:ineq_sym_core}]
Note that (\ref{eq:def_mathbb_c1}), (\ref{eq:need_verify_1}), and (\ref{eq:otherwise}) obviously hold. Moreover, (\ref{eq:need_verify_2}) is implied by (\ref{eq:moved}). Therefore, it suffices to verify (\ref{eq:need_verify_3}).




Consider an arbitrary integer $k\in \left[ {{l}^\mathbf{w}_{\alpha -1}}+1:K \right]$. We have 
\begin{align}
& \sum\limits_{{V'}\in {{\Omega }^{\left( {{l}^\mathbf{w}_{\alpha -1}} \right)}_{K,\alpha -1}}:k\in {V'}}{\sum\limits_{V\in {{\Omega }^{\left( {{l}^\mathbf{w}_{\alpha }} \right)}_{K,\alpha }}:{V'}\subseteq V}{\frac{{{c}_{V,\alpha }}}{\alpha-1 -{{l}^\mathbf{w}_{\alpha -1}}}}}
\nonumber \\
&\qquad = \sum\limits_{V\in {{\Omega }^{\left( {{l}^\mathbf{w}_{\alpha }} \right)}_{K,\alpha }}:k\in V}{\frac{\left| \left\{ {V'}\in {{\Omega }^{\left( {{l}^\mathbf{w}_{\alpha -1}} \right)}_{K,\alpha -1}}:k\in {V'}\subseteq V \right\} \right|}{\alpha-1 -{{l}^\mathbf{w}_{\alpha -1}}}{{c}_{V,\alpha }}}
\nonumber \\
&\qquad = \sum\limits_{V\in {{\Omega }^{\left( {{l}^\mathbf{w}_{\alpha }} \right)}_{K,\alpha }}:k\in V}{{{c}_{ V,\alpha }}}.
\label{eq:use_cardinality}
\end{align}
Note that
\begin{align}
& \sum\limits_{{V}'\in \Omega _{K,\alpha -1}^{\left( {{l}^\mathbf{w}_{\alpha -1}} \right)}:k\in {V}'}{ \frac{1}{{\lambda^\mathbf{w}_{\alpha }}}\sum\limits_{V\in \Omega _{K,\alpha }^{\left( {{l}^\mathbf{w}_{\alpha }} \right)}}{{{c}_{V,\alpha }}}\sum\limits_{i={{l}^\mathbf{w}_{\alpha -1}}+1}^{{{l}^\mathbf{w}_{\alpha }}}{{{c}_{\left[ 1:i \right],\alpha }}}\sum\limits_{\tau =i+1}^{\alpha }{\mathcal{I}\left\{ {V}'={{\left\langle V \right\rangle }_{\left[ 1:\alpha  \right]\backslash \left\{ \tau  \right\}}} \right\}} }
\nonumber \\
&\qquad =\frac{1}{{\lambda^\mathbf{w}_{\alpha }}}\sum\limits_{V\in \Omega _{K,\alpha }^{\left( {{l}^\mathbf{w}_{\alpha }} \right)}}{{{c}_{V,\alpha }}}\sum\limits_{i={{l}^\mathbf{w}_{\alpha -1}}+1}^{{{l}^\mathbf{w}_{\alpha }}}{{{c}_{\left[ 1:i \right],\alpha }}}\sum\limits_{\tau =i+1}^{\alpha }{\sum\limits_{{V}'\in \Omega _{K,\alpha -1}^{\left( {{l}^\mathbf{w}_{\alpha -1}} \right)}:k\in {V}'}{\mathcal{I}\left\{ {V}'={{\left\langle V \right\rangle }_{\left[ 1:\alpha  \right]\backslash \left\{ \tau  \right\}}} \right\}}}.
\label{eq:use_swap_sum}
\end{align}
Moreover,
\begin{align}
& \sum\limits_{\tau =i+1}^{\alpha } \sum\limits_{{V}'\in \Omega _{K,\alpha -1}^{\left( {{l}^\mathbf{w}_{\alpha -1}} \right)}:k\in {V}'}{\mathcal{I}\left\{ {V}'={{\left\langle V \right\rangle }_{\left[ 1:\alpha  \right]\backslash \left\{ \tau  \right\}}} \right\}}
 \nonumber \\
&=\sum\limits_{\tau =i+1}^{\alpha } \mathcal{I}\left\{ k\in {{\left\langle V \right\rangle }_{\left[ 1:\alpha  \right]\backslash \left\{ \tau  \right\}}} \right\}\nonumber\\
&=\left( \alpha-1 -i \right)\mathcal{I}\left\{ k\in V \right\}+\mathcal{I}\left\{ k\in {{\left\langle V \right\rangle }_{\left[ 1:i \right]}} \right\},\quad V\in \Omega _{K,\alpha }^{\left( {{l}^\mathbf{w}_{\alpha }} \right)},i\in \left[ {{l}^\mathbf{w}_{\alpha -1}}+1:{{l}^\mathbf{w}_{\alpha }} \right].\label{eq:sum_v'}
\end{align}
Therefore,
\begin{align}
& \sum\limits_{{V}'\in \Omega _{K,\alpha -1}^{\left( {{l}^\mathbf{w}_{\alpha -1}} \right)}:k\in {V}'}{ \frac{1}{{\lambda^\mathbf{w}_{\alpha }}}\sum\limits_{V\in \Omega _{K,\alpha }^{\left( {{l}^\mathbf{w}_{\alpha }} \right)}}{{{c}_{V,\alpha }}}\sum\limits_{i={{l}^\mathbf{w}_{\alpha -1}}+1}^{{{l}^\mathbf{w}_{\alpha }}}{{{c}_{\left[ 1:i \right],\alpha }}}\sum\limits_{\tau =i+1}^{\alpha }{\mathcal{I}\left\{ {V}'={{\left\langle V \right\rangle }_{\left[ 1:\alpha  \right]\backslash \left\{ \tau  \right\}}} \right\}} }
\nonumber \\
&  =\frac{1}{{\lambda^\mathbf{w}_{\alpha }}}\sum\limits_{V\in \Omega _{K,\alpha }^{\left( {{l}^\mathbf{w}_{\alpha }} \right)}}{{{c}_{V,\alpha }}}\sum\limits_{i={{l}^\mathbf{w}_{\alpha -1}}+1}^{{{l}^\mathbf{w}_{\alpha }}}{{{c}_{\left[ 1:i \right],\alpha }}}\left( \left( \alpha-1 -i \right)\mathcal{I}\left\{ k\in V \right\}+\mathcal{I}\left\{ k\in {{\left\langle V \right\rangle }_{\left[ 1:i \right]}} \right\} \right)
\label{eq:substi} \\
&  =\frac{1}{{\lambda^\mathbf{w}_{\alpha }}}\sum\limits_{i={{l}^\mathbf{w}_{\alpha -1}}+1}^{{{l}^\mathbf{w}_{\alpha }}}\left( \alpha-1 -i \right){{{c}_{\left[ 1:i \right],\alpha }}}\sum\limits_{V\in \Omega _{K,\alpha }^{\left( {{l}^\mathbf{w}_{\alpha }} \right)}:k\in V}{{{c}_{V,\alpha }}}
\nonumber \\
&  \quad +
\frac{1}{{\lambda^\mathbf{w}_{\alpha }}}\sum\limits_{V\in \Omega _{K,\alpha }^{\left( {{l}^\mathbf{w}_{\alpha }} \right)}:k\in V}{{{c}_{V,\alpha }}}\sum\limits_{i={{l}^\mathbf{w}_{\alpha -1}}+1}^{{{l}^\mathbf{w}_{\alpha }}}{{{c}_{\left[ 1:i \right],\alpha }}}\mathcal{I}\left\{ k\in {{\left\langle V \right\rangle }_{\left[ 1:i \right]}} \right\},
\label{eq:expand2}
\end{align}
where (\ref{eq:substi}) is obtained by substituting (\ref{eq:sum_v'}) into (\ref{eq:use_swap_sum}). 
Combining (\ref{eq:use_cardinality}) and (\ref{eq:expand2}) gives
\begin{align}
& \sum\limits_{{V}'\in \Omega _{K,\alpha -1}^{\left( {{l}^\mathbf{w}_{\alpha -1}} \right)}:k\in {V}'}{{{c}_{{V}',\alpha -1}}}
\nonumber \\
& =\sum\limits_{V\in \Omega _{K,\alpha }^{\left( {{l}^\mathbf{w}_{\alpha }} \right)}:k\in V}{{{c}_{V,\alpha }}}+\frac{1}{{\lambda^\mathbf{w}_{\alpha }}}\sum\limits_{V\in \Omega _{K,\alpha }^{\left( {{l}^\mathbf{w}_{\alpha }} \right)}:k\in V}{{{c}_{V,\alpha }}}\sum\limits_{i={{l}^\mathbf{w}_{\alpha -1}}+1}^{{{l}^\mathbf{w}_{\alpha }}}{{{c}_{\left[ 1:i \right],\alpha }}}\mathcal{I}\left\{ k\in {{\left\langle V \right\rangle }_{\left[ 1:i \right]}} \right\}.
\label{eq:before_cases}
\end{align}
Now consider the following two cases.

\emph{(Case 1)}
$k\in \left[ {{l}^\mathbf{w}_{\alpha -1}}+1:{{l}^\mathbf{w}_{\alpha }} \right]$: We have
\begin{align}
\sum\limits_{V\in \Omega _{K,\alpha }^{\left( {{l}^\mathbf{w}_{\alpha }} \right)}:k\in V}{{{c}_{V,\alpha }}}&=\sum\limits_{V\in \Omega _{K,\alpha }^{\left( {{l}^\mathbf{w}_{\alpha }} \right)}}{{{c}_{V,\alpha }}}\nonumber\\
&={\lambda^\mathbf{w}_{\alpha }}\label{eq:invokesomelem}
\end{align}
and
\begin{align}
\mathcal{I}\left\{ k\in {{\left\langle V \right\rangle }_{\left[ 1:i \right]}} \right\}=\mathcal{I}\left\{ k\in \left[ 1:i \right] \right\},\quad V\in \Omega _{K,\alpha }^{\left( {{l}^\mathbf{w}_{\alpha }} \right)}, i\in \left[ {{l}^\mathbf{w}_{\alpha -1}}+1:{{l}^\mathbf{w}_{\alpha }} \right],\label{eq:changeset}
\end{align}
where (\ref{eq:invokesomelem}) is due to (\ref{eq:partial_sum}). 
Continuing from (\ref{eq:before_cases}),
\begin{align}
& \sum\limits_{{V}'\in \Omega _{K,\alpha -1}^{\left( {{l}^\mathbf{w}_{\alpha -1}} \right)}:k\in {V}'}{{{c}_{{V}',\alpha -1}}} \nonumber\\
&=\sum\limits_{V\in \Omega _{K,\alpha }^{\left( {{l}^\mathbf{w}_{\alpha }} \right)}:k\in V}{{{c}_{V,\alpha }}}+\frac{1}{{\lambda^\mathbf{w}_{\alpha }}}\sum\limits_{V\in \Omega _{K,\alpha }^{\left( {{l}^\mathbf{w}_{\alpha }} \right)}:k\in V}{{{c}_{V,\alpha }}}\sum\limits_{i={{l}^\mathbf{w}_{\alpha -1}}+1}^{{{l}^\mathbf{w}_{\alpha }}}{{{c}_{\left[ 1:i \right],\alpha }}}\mathcal{I}\left\{ k\in \left[ 1:i \right] \right\}\label{eq:duetotwo}\\
&=\sum\limits_{V\in \Omega _{K,\alpha }^{\left( {{l}^\mathbf{w}_{\alpha }} \right)}:k\in V}{{{c}_{V,\alpha }}}+\frac{1}{{\lambda^\mathbf{w}_{\alpha }}}\sum\limits_{i=k}^{{{l}^\mathbf{w}_{\alpha }}}{{{c}_{\left[ 1:i \right],\alpha }}}\sum\limits_{V\in \Omega _{K,\alpha }^{\left( {{l}^\mathbf{w}_{\alpha }} \right)}:k\in V}{{{c}_{V,\alpha }}}\nonumber\\
&=  {\lambda^\mathbf{w}_{\alpha }}+\sum\limits_{i=k}^{{{l}^\mathbf{w}_{\alpha }}}{{{c}_{\left[ 1:i \right],\alpha }}}\label{eq:duetofirst} \\
& ={{w}_{k}},\nonumber
\end{align}
where (\ref{eq:duetotwo}) and (\ref{eq:duetofirst}) are due to (\ref{eq:changeset}) and (\ref{eq:invokesomelem}), respectively.

\emph{(Case 2)}
$k\in \left[ {{l}^\mathbf{w}_{\alpha }}+1:K \right]$: We have
\begin{align}
 \sum\limits_{{V}'\in \Omega _{K,\alpha -1}^{\left( {{l}^\mathbf{w}_{\alpha -1}} \right)}:k\in {V}'}{{{c}_{{V}',\alpha -1}}}& \quad =\sum\limits_{V\in \Omega _{K,\alpha }^{\left( {{l}^\mathbf{w}_{\alpha }} \right)}:k\in V}{{{c}_{V,\alpha }}} \label{eq:iszero}\\
& \quad ={{w}_{k}},\nonumber
\end{align}
where (\ref{eq:iszero}) follows by (\ref{eq:before_cases}) and the fact that
\begin{align*}
\mathcal{I}\left\{ k\in {{\left\langle V \right\rangle }_{\left[ 1:i \right]}} \right\}=0,\quad V\in \Omega _{K,\alpha }^{\left( {{l}^\mathbf{w}_{\alpha }} \right)}, i\in \left[ {{l}^\mathbf{w}_{\alpha -1}}+1:{{l}^\mathbf{w}_{\alpha }} \right].
\end{align*}
This completes the verification of (\ref{eq:need_verify_3}).
\end{IEEEproof}

\begin{IEEEproof}[Proof of Part (2) of Lemma~\ref{lem:ineq_sym_core}]
Note that 
\begin{align}
& {\sum\limits_{\tau =i+1}^{\left|V\right| }{H\left( {{X}_{{{\left\langle V \right\rangle }_{\left[ 1:\left|V\right|  \right]\backslash \left\{ \tau  \right\}}}}} \right)}} \nonumber\\
&={\sum\limits_{\tau =i+1}^{\left|V\right| }{H\left( {{X}_{{{\left\langle V \right\rangle }_{\left[ 1:\left|V\right|  \right]\backslash \left\{ \tau  \right\}}}}}| {{X}_{{{\left\langle V \right\rangle }_{\left[ 1:i \right]}}}}  \right)}} +{\sum\limits_{\tau =i+1}^{\left|V\right|}{H\left({{X}_{{{\left\langle V \right\rangle }_{\left[ 1:i \right]}}}} \right)}}\nonumber\\
&={\sum\limits_{\tau =i+1}^{\left|V\right| }{H\left( {{X}_{{{\left\langle V \right\rangle }_{\left[ 1:\left|V\right|  \right]\backslash \left\{ \tau  \right\}}}}}| {{X}_{{{\left\langle V \right\rangle }_{\left[ 1:i \right]}}}}  \right)}} +\left(\left|V\right|-i\right){H\left({{X}_{{{\left\langle V \right\rangle }_{\left[ 1:i \right]}}}} \right)}\nonumber\\
&\geq\left(\left|V\right|-1-i\right){H\left( {{X}_{V}}| {{X}_{{{\left\langle V \right\rangle }_{\left[ 1:i \right]}}}}  \right)} +\left(\left|V\right|-i\right){H\left({{X}_{{{\left\langle V \right\rangle }_{\left[ 1:i \right]}}}} \right)}\label{eq:invokeHan1}\\
& =\left(\left|V\right|-1-i\right){H\left( {{X}_{V}}  \right)} +{H\left({{X}_{{{\left\langle V \right\rangle }_{\left[ 1:i \right]}}}} \right)},\quad i\in \left[ 0:\left|V\right| -1 \right],\label{eq:extended}
\end{align}
where (\ref{eq:invokeHan1}) is due to Han's inequality \cite{han1978}. 
We have
\begin{align}
& \sum\limits_{{V}'\in \Omega _{K,\alpha -1}^{\left( {{l}^\mathbf{w}_{\alpha -1}} \right)}}\frac{{\theta^\mathbf{w}_{\alpha }}}{{\lambda^\mathbf{w}_{\alpha }}} \sum\limits_{V\in \Omega _{K,\alpha }^{\left( {{l}^\mathbf{w}_{\alpha }} \right)}:{V}'\subseteq V}{{c}_{V,\alpha }}H\left( {{X}_{{{V}'}}} \right)
\nonumber \\
&=\frac{{\theta^\mathbf{w}_{\alpha }}}{{\lambda^\mathbf{w}_{\alpha }}}\sum\limits_{V\in \Omega _{K,\alpha }^{\left( {{l}^\mathbf{w}_{\alpha }} \right)}}{{c}_{V,\alpha }}\sum\limits_{{V}'\in \Omega _{K,\alpha -1}^{\left( {{l}^\mathbf{w}_{\alpha -1}} \right)}:{V}'\subseteq V}    H\left( {{X}_{{{V}'}}} \right)\nonumber\\
&=\frac{{\theta^\mathbf{w}_{\alpha }}}{{\lambda^\mathbf{w}_{\alpha }}}\sum\limits_{V\in \Omega _{K,\alpha }^{\left( {{l}^\mathbf{w}_{\alpha }} \right)}}{{c}_{V,\alpha }}{\sum\limits_{\tau ={{l}^\mathbf{w}_{\alpha -1}}+1}^{\alpha }{H\left( {{X}_{{{\left\langle V \right\rangle }_{\left[ 1:\alpha  \right]\backslash \left\{ \tau  \right\}}}}} \right)}} \nonumber\\
&\ge\frac{{\theta^\mathbf{w}_{\alpha }}}{{\lambda^\mathbf{w}_{\alpha }}}\sum\limits_{V\in \Omega _{K,\alpha }^{\left( {{l}^\mathbf{w}_{\alpha }} \right)}}{{c}_{V,\alpha }}\left(\left( \alpha-1 -{{l}^\mathbf{w}_{\alpha -1}} \right)H\left( {{X}_{V}} \right)+H\left( {{X}_{\left[ 1:{{l}^\mathbf{w}_{\alpha -1}} \right]}} \right)\right)\label{eq:hand1}\\
& = \sum\limits_{V\in {{\Omega }^{\left( {{l}^\mathbf{w}_{\alpha }} \right)}_{K,\alpha }}}{{{c}_{V,\alpha }}H\left( {{X}_{V}} \right)}
- \frac{1}{{\lambda^\mathbf{w}_{\alpha }}}\sum\limits_{i={{l}^\mathbf{w}_{\alpha -1}}+1}^{{{l}^\mathbf{w}_{\alpha }}}{\left( \alpha -i-1 \right){{c}_{\left[ 1:i \right],\alpha }}}\sum\limits_{V\in {{\Omega }^{\left( {{l}^\mathbf{w}_{\alpha }} \right)}_{K,\alpha }}}{{{c}_{V,\alpha }}H\left( {{X_V}} \right)}\nonumber\\
&\quad+\left({{c}_{\left[ 1:l_{\alpha -1}^{\mathbf{w}}  \right],\alpha }}-{{c}_{\left[ 1:l_{\alpha -1}^{\mathbf{w}}  \right],\alpha -1}}\right)H\left( {{X}_{\left[ 1:{{l}^\mathbf{w}_{\alpha -1}} \right]}} \right),\label{eq:invoketwoid}
\end{align}
where (\ref{eq:hand1}) follows by (\ref{eq:extended}), and (\ref{eq:invoketwoid}) is due to (\ref{eq:partial_sum}) and (\ref{eq:moved}).
Moreover,
\begin{align}
& \sum\limits_{{V'}\in {{\Omega }^{\left( {{l}^\mathbf{w}_{\alpha -1}} \right)}_{K,\alpha -1}}}\frac{1}{{\lambda^\mathbf{w}_{\alpha }}}\sum\limits_{V\in {{\Omega }^{\left( {{l}^\mathbf{w}_{\alpha }} \right)}_{K,\alpha }}}{{{c}_{V,\alpha }}\sum\limits_{i={{l}^\mathbf{w}_{\alpha -1}}+1}^{{{l}^\mathbf{w}_{\alpha }}}{{{c}_{\left[ 1:i \right],\alpha }}{\sum\limits_{\tau =i\text{+}1}^{\alpha }{\mathcal{I}\left\{ {V'}={{\left\langle V \right\rangle }_{\left[ 1:\alpha  \right]\backslash \left\{ \tau  \right\}}} \right\}H\left( {{X}_{V'}} \right)}}}}
\nonumber \\
&= \frac{1}{{\lambda^\mathbf{w}_{\alpha }}}\sum\limits_{V\in {{\Omega }^{\left( {{l}^\mathbf{w}_{\alpha }} \right)}_{K,\alpha }}}{{{c}_{V,\alpha }}\sum\limits_{i={{l}^\mathbf{w}_{\alpha -1}}+1}^{{{l}^\mathbf{w}_{\alpha }}}{{{c}_{\left[ 1:i \right],\alpha }}{\sum\limits_{\tau =i\text{+}1}^{\alpha }{H\left( {{X}_{{{\left\langle V \right\rangle }_{\left[ 1:\alpha  \right]\backslash \left\{ \tau  \right\}}}}} \right)}}}}
\nonumber \\
& \ge \frac{1}{{\lambda^\mathbf{w}_{\alpha }}}\sum\limits_{V\in {{\Omega }^{\left( {{l}^\mathbf{w}_{\alpha }} \right)}_{K,\alpha }}}{{{c}_{V,\alpha }}\sum\limits_{i={{l}^\mathbf{w}_{\alpha -1}}+1}^{{{l}^\mathbf{w}_{\alpha }}}{{{c}_{\left[ 1:i \right],\alpha }}\left( \left( \alpha-1 -i \right)H\left( {{X}_{V}} \right)+H\left( {{X}_{\left[ 1:i \right]}} \right) \right)}}
\label{eq:invokeextended}\\
& = \frac{1}{{\lambda^\mathbf{w}_{\alpha }}}\sum\limits_{i={{l}^\mathbf{w}_{\alpha -1}}+1}^{{{l}^\mathbf{w}_{\alpha }}}{\left( \alpha -1-i \right){{c}_{\left[ 1:i \right],\alpha }}}\sum\limits_{V\in {{\Omega }^{\left( {{l}^\mathbf{w}_{\alpha }} \right)}_{K,\alpha }}}{{{c}_{V,\alpha }}H\left( {{X}_{V}} \right)}\nonumber\\
&\quad+\frac{1}{{\lambda^\mathbf{w}_{\alpha }}}\sum\limits_{i={{l}^\mathbf{w}_{\alpha -1}}+1}^{{{l}^\mathbf{w}_{\alpha }}}{{{c}_{\left[ 1:i \right],\alpha }}H\left( {{X}_{\left[ 1:i \right]}} \right)}\sum\limits_{V\in {{\Omega }^{\left( {{l}^\mathbf{w}_{\alpha }} \right)}_{K,\alpha }}}{{{c}_{V,\alpha }}}\nonumber\\
&=\frac{1}{{\lambda^\mathbf{w}_{\alpha }}}\sum\limits_{i={{l}^\mathbf{w}_{\alpha -1}}+1}^{{{l}^\mathbf{w}_{\alpha }}}{\left( \alpha-1 -i \right){{c}_{\left[ 1:i \right],\alpha }}}\sum\limits_{V\in {{\Omega }^{\left( {{l}^\mathbf{w}_{\alpha }} \right)}_{K,\alpha }}}{{{c}_{V,\alpha }}H\left( {{X}_{V}} \right)}+\sum\limits_{i={{l}^\mathbf{w}_{\alpha -1}}+1}^{{{l}^\mathbf{w}_{\alpha }}}{{{c}_{\left[ 1:i \right],\alpha }}H\left( {{X}_{\left[ 1:i \right]}} \right)},
\label{eq:hand2}
\end{align}
where (\ref{eq:invokeextended}) and (\ref{eq:hand2}) are due to (\ref{eq:extended}) and (\ref{eq:partial_sum}), respectively. 
Now one can readily complete the proof by combining (\ref{eq:hand1}) and (\ref{eq:hand2}) and invoking the fact that ${{c}_{\left[ 1:i \right],\alpha-1}}={{c}_{\left[ 1:i \right],\alpha}}$ when  $i\in\left[1: l_{\alpha -1}^{\mathbf{w}}-1\right]$.
\end{IEEEproof}

\bibliographystyle{./IEEEtran}
\end{document}